\documentclass[letter,11pt]{jhep3}
\usepackage{amsmath,amssymb,enumerate,epsfig}

\newcommand{\bignone}{}
\newcommand{\emdash}{---}
\newcommand{\mathd}{\mathrm{d}}
\newcommand{\mathi}{\mathrm{i}}
\newcommand{\mathpi}{\pi}
\newcommand{\tmem}[1]{{\em #1\/}}
\newcommand{\tmop}[1]{\ensuremath{\operatorname{#1}}}
\newcommand{\tmstrong}[1]{\textbf{#1}}

\newenvironment{enumeratenumeric}{\begin{enumerate}[1.] }{\end{enumerate}}
\newenvironment{itemizedot}{\begin{itemize} }{\end{itemize}}

\title{The fifth dimension as an analogue computer for strong interactions at
the LHC}

\author{Johannes Hirn\\ Department of Physics, Yale University, New Haven, CT
  06520\\ \email{johannes.hirn@yale.edu}}
\author{Ver\'onica Sanz\\ Department of Physics, Boston University, Boston, MA
  02215\\
\email{vsanz@bu.edu}}

\abstract{We present a mechanism to get $S \simeq 0$ or even negative, without
  bringing into play the SM fermion sector. This mechanism can be applied to a
  wide range of 5D models, including composite Higgs and Higgsless models. As
  a realization of the mechanism we introduce a simple model, although the effect on $S$ does not rely on the underlying dynamics
  generating the background. Models that include this mechanism enjoy the
  following features: weakly-coupled light resonances (as light as 600 GeV)
  and degenerate or inverted resonance spectrum.}

\begin{document}

\section{The analogue computer in four steps}\label{analogueintro}

Before we give a more formal Introduction in Section 2, we briefly describe
the purpose of this paper, as summarized by the four steps of Fig.\ref{purpose}.

\EPSFIGURE[!h]{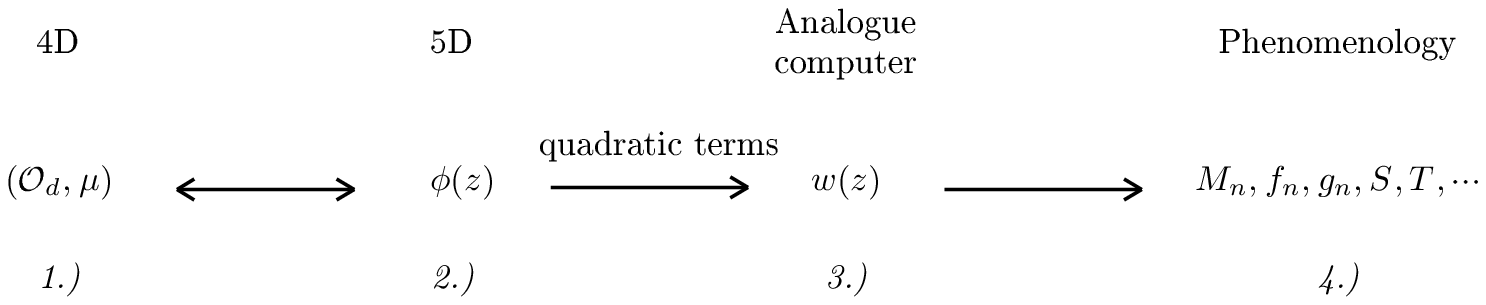}{\label{purpose}The role of the the
  analogue computer.}

{\tmem{1.) }}Assume that a strongly-coupled sector is (partly or completely)
responsible of electroweak symmetry breaking (EWSB). The strong sector can be
described by, for example, physical resonances (mesons) and, in general, by a
set of operators with couplings to the EW sector: \{$\mathcal{O}_d$\}. $d$
denotes the scaling dimension of the operator and $\langle \mathcal{O}_d
\rangle$ its vev.

{\tmem{2.) }}Since computations in the strong sector are difficult,
unreliable or impossible, one can use a tool, the fifth dimension, to extract
physical quantities. The procedure is the so-called {\tmem{holographic
recipe}} {\footnote{See
{\cite{Maldacena:1997re,Gubser:1998bc,Witten:1998qj,Arkani-Hamed:2000ds,Pomarol:2000hp}}
and references therein for more details on the correspondence conjecture.}}:
the properties of the set $\{\mathcal{O}_d \}$ \ are substituted by a set of
bulk fields $\{\phi (z)\}$. This is achieved by relating the mass of the field
to the dimension of the operator by $m_{\phi}^2 = d (d - 4)$ in units of the
curvature scale, the running scale $\mu$ to $1 / z$, the vev of $\langle
\mathcal{O}_d \rangle$ to that of $\langle \phi \rangle$, and transformation
properties of $\mathcal{O}_d$ to the ones of $\phi$...

The degree of reliability of this procedure depends on the 5D gauge coupling,
or in other words, on the corresponding 4D large-$N$ expansion.

At this level, extracting physical quantities depends on one's ability to
choose the right set $\{\phi\}$ and to solve coupled equations of motion. As
you can imagine, this task is also rather lengthy, specially when several bulk
fields are relevant for the discussion.

{\tmem{3.)}} We propose to go a step further by realizing that, {\tmem{at the
quadratic level,}} {\tmem{the effect of background fields on the resonances
and Goldstone bosons is equivalent to introducing an {\tmstrong{effective
metric}} and modifying the boundary conditions (BCs).}} Even when the
background fields produce light modes {\cite{Erlich:2005qh,daRold:2005zs}},
one can still perform this rewriting while keeping the light fields in the
spectrum. This is our {\tmem{analogue computer}}: whatever the background
fields $\{\phi\}$, they result in a particular form of the effective metric
felt by the mesons.

This procedure, valid only at the quadratic level, is more relevant than it
seems: except in particular cases {\cite{Katz:2005ir}}, quadratic interactions
are the only reliable quantities one can compute in these kind of models
{\footnote{We'd like to thank Ami Katz for illuminating discussions on this
point.}}.

{\tmem{4.)}} Once the analogue computer is built, the rest is much easier: the
scale of EWSB, the spectrum, decay constants, the couplings to the SM fermions
and, in particular, the electroweak precision parameters like $S$ and $T$ can
be computed straight away just in terms of a metric with few coefficients.

The effective metric has few coefficients because the effect of these
background fields on observables decreases with the dimension $d$. In
particular, only two condensates are relevant for the discussion, and $S$ can
be correlated with the properties of the strong sector mesons.

The paper is organized as follows. In Section 2, we discuss the situation of
$S$-parameter studies both from the 4D and 5D side. In Section 3, we introduce
the relevant parameters in our effective model. Section 4 goes into some
details of the effect of background fields on $S$, and how the same physics is
described by our analogue computer. Section 5 gives the result for $S$ in the
parameter space we consider. Section 6 describes the consequences on the
spectrum. Section 7 presents a purely 4D interpretation of the result. We
present our conclusions in Section 8. Various Appendices give details on
technical points.

\section{What holography has to say about the $S$ parameter}\label{intro}

Electroweak precision tests (EWPTs) seem to tell us that new physics models
have to follow very determined patterns, as is the case for the SM itself. For
example, one can suppress contributions to $\Delta \rho$ by enforcing
custodial symmetry in the new sector {\cite{Sikivie:1980hm}}. This symmetry
can be embedded into a larger symmetry that protects new couplings, like
anomalous contributions to the $Z \rightarrow b \bar{b}$
{\cite{Agashe:2006at}}.

\subsection{A short account of the problem}

We focus on the $S$ parameter here, even though technicolor
{\cite{Weinberg:1979bn,Susskind:1978ms}} is also known to encounter other
difficulties once it is extended to include fermion masses
{\cite{Eichten:1979ah,Dimopoulos:1979es}}. In general, the new physics
contributions to \ $S$ are of order
\begin{eqnarray}
  S_{\tmop{tree}} & \sim & \frac{N}{4 \mathpi}  \left( \frac{v}{f} \right)^2, 
  \label{Sest2}
\end{eqnarray}
where $N$ is a measure of the size of the new sector {\cite{Peskin:1991sw}}:
it is an effective number of degrees of freedom. In (\ref{Sest2}), $f / v$
represents the little hierarchy between the decay constant $f$ of the three
Goldstone bosons (GBs) that are eaten by the $W$'s and $Z$ and the vev $v$ of
the composite Higgs (built up of pseudo-GBs of a larger group). In all known
models, $\left( f / v \right)^2$ cannot be made much larger than a factor
twenty {\cite{hep-ph/0306259,Agashe:2004rs}}. At the other extreme is the minimal case
without a light Higgs. This is also the most disfavorable one: resonance and
symmetry-breaking (SB) scales are then tied together, so that the formula
(\ref{Sest2}) can be used with replacement $v \rightarrow f$.

As for $N$, it can be interpreted more specifically {\tmem{1.)}} for a 4D
strongly-interacting theory, as the number of colors or {\tmem{2.)}} for a 5D
model, as the number of Kaluza-Klein (KK) states that can be generated at low
energies before strong coupling sets in. The relation between these two
quantities is discussed in Section \ref{relevant}.

The point is that, in either case, a large $N$ is what makes the description
in terms of resonances/KKs perturbative. Thus, in order for such scenarios to
remain predictive at the few TeV scale, we need either a physical Higgs boson
or a large $N$ to ensure perturbativity. The first possibility can result in a
small $S$: if there is a (composite) Higgs, the first resonance may be heavy
enough as not to produce a large $S$, see {\cite{Contino:2006qr}} for the
latest update. On the other hand, if the job of keeping $W W$ scattering
perturbative is to be done by the resonances alone, $N$ cannot be too small,
and one expects the lightest resonances below 2 TeV. We focus on this less
favorable case here: the mechanism may then be applied to the more favorable
situation with a composite Higgs.

The basic problem is that strong interactions readily produce large deviations
from the SM. Among all the relevant electroweak parameters {\cite{Barbieri:2004qk,hep-ph/0604111}}, the main culprit is $S$. 
Whereas we know experimentally that $- 0.4 \lesssim S \lesssim
0.2$ at the $3 \sigma$ level {\cite{Yao:2006px}}, this bound is easily
exceeded in scenarios of dynamical EWSB. Beyond using equation (\ref{Sest2}),
possible methods of estimating $S$ for strong dynamics include using QCD as an
analogue computer {\cite{Holdom:1990tc,Peskin:1991sw}}, combined with our best
handle on strong interactions, namely the large-$N$ limit. This yields $S \sim
0.1 N$, so that a version of {\tmem{technicolor built as a simple rescaled QCD
is experimentally excluded}}. Since only low-$N$ theories would stand a chance
of passing the constraint, computing $S$ would be hopeless.

Such a picture was corroborated using a 5D approach in {\cite{Luty:2004ye}}.
The 5D models in question are constructed to embody the same physics
(confinement, symmetry-breaking) in a dual description in terms of mesons. For
them to remain perturbative above the first few KK resonance (to be understood
as the techni-mesons), the 5D gauge coupling should remain small in units of
the AdS curvature. Since this quantity directly corresponds to the $1 / N$ of
a 4D theory, the 5D description fails for the same reason as the 4D one
{\cite{Barbieri:2003pr}}. This is generically valid
{\cite{Csaki:2003zu,hep-ph/0309189,daRold:2005zs}}, unless the profile of the SM
fermions is chosen to be nearly flat
{\cite{Agashe:2003zs}}.

\subsection{Our solution}\label{our-sol}

In the present paper, we draw upon the 5D approach, without invoking
cancellations with the fermion sector. In that approach, it is the bulk dynamics
that generate large contributions to the $S$ parameter (proportional to $N$).
However, since there are two competing contributions with different signs
{\emdash}respectively from the vector/axial resonances{\emdash} there is no
generic value for $S$ in a strongly-interacting model. In fact, we find that
there is a significant fraction of parameter space for which $S$ passes the
experimental constraint.

The added ingredient compared to previous 5D modeling comes from holographic
QCD {\cite{Son:2003et,Erlich:2005qh,daRold:2005zs}}. Namely, we refine the 5D
model by matching with the first terms in an OPE of the two-point functions
{\cite{Hirn:2005vk}}. In {\cite{Hirn:2006nt}}, we considered matching the 5D
model to a different high-energy behavior than that of QCD: we called this
Holographic Technicolor. Here, we go further: we present a {\tmem{general
parametrization}} of terms quadratic in spin-1 resonances. This allows us to
correlate the experimental value of $S$ with properties of the new physics
sector. The result thus does not depend on the details of the underlying 5D
modeling. Still, we provide as an example a model consistent with gravity,
which implies definite signs and magnitudes for the parameters of our analogue
computer. These allow for $S \lesssim 0$.

This parametrization serves as an {\tmem{analogue computer}} for strong
interactions, and can be applied to 5D models with a physical Higgs scalar,
such as composite Higgs models {\cite{Agashe:2004rs}} or gaugephobic Higgs
{\cite{Cacciapaglia:2006mz}}. Note that, once such a model passes the
constraint on $S$, the remaining experimental constraints on resonance masses
come mainly from direct production. For numerical applications, we consider \
the extreme case where the lightest resonance has a mass of $600 \tmop{GeV}$.
It will turn out that a vanishing or slightly negative $S$ is correlated with
a degenerate spectrum, or even with an inverted spectrum, so the lightest
resonance is a techni-$a_1$ rather than a techni-$\tmop{rho}$.

This study leads us to make the following claims. {\tmem{1.)}} As was the case
for 4D strong dynamics, the value of $S$ cannot be {\tmem{predicted}} in
general for 5D models.{\tmem{ 2.)}} $S$ can change sign in a {\tmem{weakly
coupled}} 5D model {\cite{Hirn:2006nt}}. {\tmem{3.)}} Setting the value of $S$
to be within the experimental bounds, one finds correlations between the
spectrum, the couplings, the OPE and the scale of electroweak symmetry
breaking.

Point {\tmem{1.)}} is a known fact: although the natural estimate for $S$ with
light resonances is positive and order one, one can always rescue the
particular model by switching on some compensating effects. This job becomes
harder as the resonances are more weakly coupled (large-$N$), but was shown to
be feasible in {\cite{Hirn:2006nt}} and is further discussed here. Point
{\tmem{2.)}} is new in the sense that we are dealing with a weakly-coupled and
light sector of resonances coupled to EWSB, and still we can reduce the value
of $S$ to be within experimental limits and even change its sign {\footnote{We
stress again that this occurs without resorting to cancellations with the
fermion sector {\cite{Cacciapaglia:2004jz}}.}}. Point {\tmem{3.)}} is the
subject of this paper: the use of the 5th dimension as a tool is very powerful
to describe these correlations in a calculable way. In fact, we find that, in
the simplest toy model for bulk fields one could write down, the cumulative
effect of bulk dynamics can indeed go in the direction of lowering $S$, and go
so far as to make it negative for reasonable values of the parameters.

As mentioned in Section \ref{intro}, the present paper discusses a class of
model in which $S_{\tmop{tree}}$ may vanish or even become negative. The
second possibility is even more welcome for the following reason. The value of
$S$ is obtained by taking the difference between a model and the SM, used at
loop level. The SM Higgs reference mass thus enters the calculation. The tree
level contribution of (\ref{Sest2}) has to be corrected by the running loop
effects {\cite{Holdom:1990tc,Peskin:1991sw}}. One can estimate these effects
by running until the scale of new physics $\Lambda$
\begin{eqnarray}
  S & = & \left. S_{\tmop{tree}} + \frac{1}{12 \mathpi}  \left( \ln \left(
  \frac{\Lambda^2}{m_H^2} \right) - \frac{1}{6} \right) \right) . 
  \label{Sloop}
\end{eqnarray}
This effect is sizable and of order 0.1 for $\Lambda \sim 1 \tmop{TeV}$, so
that we will require $- 0.5 < S_{\tmop{tree}} < 0.1$. Therefore, rather than
considering the bound $- 0.4 \lesssim S \lesssim 0.2$ on $S$, we use a bound
$- 0.5 \lesssim S_{\tmop{tree}} \lesssim 0.1$ and omit the subscript ``tree''.

\section{Some useful definitions}\label{5Dmodel}

Our analogue computer is a 5D model, where the KK modes can also be
interpreted as the resonances of a strongly-interacting 4D theory. On the 5D
side, we assume a conformally flat metric
\begin{eqnarray}
  \mathd s^2 & = & w \left( z \right)^2  \left( \eta_{\mu \nu} \mathd x^{\mu}
  \mathd x^{\nu} - \mathd z^2 \right),  \label{2.2}
\end{eqnarray}
where $z$ is the extra coordinate, defined on an interval $l_0 \leqslant z
\leqslant l_1$. Appropriate BCs will be enforced at the endpoints $l_0$ (the
{\tmem{UV brane}}) and $l_1$ (the {\tmem{IR brane}}). $w (z)$ is the warp
factor: $w (z) = 1, l_0 / z$ corresponds to flat space and AdS respectively.

For applications to Holographic Technicolor, the interesting metrics are the
so-called {\tmem{gap-metrics}} {\cite{Hirn:2006nt}}, which decrease away from
the UV as AdS or faster. This warping of the metric is ultimately responsible
of the existence of two sectors in the spectrum: the {\tmem{ultra-light}}
({\tmem{UL}}) sector consisting of the SM fields $W, Z, \gamma$  and the {\tmem{Kaluza-Klein-sector}} ({\tmem{KK}}). The
gap between them will be denoted in general as $G$.

We will use metrics that are asymptotically AdS on the UV boundary, and break
conformal invariance near the IR
\begin{eqnarray}
  w (z) & = & \frac{l_0}{z} f \left( \frac{z}{l_1} \right),  \label{met1}
\end{eqnarray}
where $f (0) = 1$ {\footnote{In general, we ultimately use $l_0 \ll l_1$ for
numerical applications. Therefore, for all practical purposes, it does not
matter whether one imposes $f \left( 0 \right) = 1$ or $f \left( l_0 / l_1
\right) = 1$.}}. In most of the paper we will consider a simple
parametrization of deviations from AdS
\begin{eqnarray}
  f \left( \frac{z}{l_1} \right) & = & \exp \left( \frac{o_{V, A}}{2 d (d -
  1)} \left( \frac{z}{l_1} \right)^{2 d} \right) .  \label{metric}
  \label{metric2}
\end{eqnarray}
This can be obtained effectively by adding a $L R$ kinetic term in the bulk with
an appropriate profile, as in {\cite{Hirn:2006nt}}. Section \ref{point1} and
Appendix \ref{appscalar} explain how two different effective metrics can be
generated in a 5D model. Note that the phenomenology is not very sensitive to
the particular form of $f (z)$ in the IR.

In the present paper, fermions are localized on the UV brane for simplicity.
Therefore, the $S$ parameter we compute here is a pure gauge contribution. In
this way, flavor issues can be addressed separately from constraints on the
$S$.

Now let us consider a $\mathcal{G} \supset \tmop{SU} \left( 2 \right)_L \times
\tmop{SU} \left( 2 \right)_R \times \mathrm{U} \left( 1 \right)_{B - L}$ bulk
gauge symmetry. The $L R$ symmetry is necessary for custodial symmetry
{\cite{Agashe:2003zs}}. It is also included in the $O (3)$ that suppresses
deviations from the SM in $Z \rightarrow b \bar{b}$ {\cite{Agashe:2006at}}.
The action is invariant under ``parity'' $L \leftrightarrow R$. We will denote
the common $\tmop{SU} \left( 2 \right)_L \times \tmop{SU} \left( 2 \right)_R$
gauge coupling by $g_5^2$, which has dimensions of length. The ratio between
this and the $\mathrm{U} \left( 1 \right)_{B - L}$ coupling $\widetilde{g_5}$
can then be chosen in order to reproduce the experimental $M_Z / M_W$. We will
not need $\widetilde{g_5}$ further in the present paper.

The breaking patterns that are relevant for phenomenology can be summarized as
follows:
\begin{itemizedot}
  \item $\tmop{SU} (N_f)_L \times \tmop{SU} (N_f)_R \rightarrow \tmop{SU}
  (N_f)_V$ \ near/on the IR.
  
  \item $\mathcal{G} \rightarrow \varnothing \tmop{or} U (1)_Y \times
  \tmop{SU} (2)_L $ on the UV brane for {\emdash}respectively{\emdash}
  Holographic QCD or the EW case.
\end{itemizedot}
\subsection{Relevant parameters}\label{relevant}

Given the setup of the previous section, the size of deviations from SM
Physics can be estimated by knowing:
\begin{itemizedot}
  \item {\tmem{1.)}} The gap between the UL and KK sectors : $G \propto \left(
  \frac{M_{\tmop{KK}}}{M_W} \right)^2$
  
  \item {\tmem{2.)}} The size of the KK sector contributing to the EWSB
  sector: $N_{\tmop{KK}} \propto \left( \frac{4 \pi f}{M_{\tmop{KK}}}
  \right)^2$ \ 
\end{itemizedot}
We discuss these in turn.

{\tmem{1.)}} In a 5D model, $G$ depends on the warping of space-time. If the
only source of EWSB is via boundary conditions on the IR brane (Higgsless
models) the value for $G$ is simply given by a geometrical factor: $G$ is just
a number in flat space whereas it is a parametrically large factor
{\emdash}$\log (l_1 / l_0)${\emdash} for pure AdS. Large localized kinetic
terms can increase $G$. For example, one can modify the spectrum in flat space
by adding large localized kinetic terms in the $\tmop{IR}$ brane, effectively
mimicking a warp factor.

{\tmem{2.)}} Using NDA in 5D, one can show
{\cite{Manohar:1983md,Georgi:1986kr,Luty:1997fk,Cohen:1997rt,Chacko:1999hg}}
that the loop expansion for a 5D gauge field theory breaks down around the
scale
\begin{eqnarray}
  \Lambda_{\tmop{UV}} & = & \frac{24 \pi^3}{g_5^2} .  \label{LUV}
\end{eqnarray}
Using the standard definition for $N$
\begin{eqnarray}
  \frac{l_0}{g_5^2} & \equiv & \frac{N}{12 \pi^2},  \label{NKK}
\end{eqnarray}
which matches 4D and 5D correlators in the large energy limit, we can write
\begin{eqnarray}
  \Lambda_{\tmop{UV}} l_0 & = & 2 \pi N .  \label{LN}
\end{eqnarray}
In this language, large-$N$ (4D) expansion corresponds to weak coupling (5D).

However, beyond the AdS case, the 5D expansion parameter $N$ given by
(\ref{NKK}) does not coincide with the size of the low energy sector,
$N_{\tmop{KK}}$. One has to realize that the result (\ref{LUV}) holds for
processes that would be localized on the UV brane, where the warp factor is
normalized to one. For metrics as in Eq.(\ref{met1}), experiments carried out
on the UV brane (where fermions are located) have a typical cutoff $2 \pi N /
l_0 \gg 2 \pi N / l_1$. The result (\ref{LUV}) gets redshifted for processes
localized at a position $z_{\ast}$
{\cite{Pomarol:2000hp,Randall:2001gc,Randall:2002tg}}, i.e. if the involved
overlap integrals are dominated by contributions around $z_{\ast}$. The scale
at which a process localized in $z_{\ast}$ becomes strongly-coupled is thus
\begin{eqnarray}
  \Lambda \left( z_{\ast} \right) & = & \frac{24 \mathpi^3}{g_5^2} w \left(
  z_{\ast} \right) .  \label{Lambdaz}
\end{eqnarray}
For metrics of the form (\ref{met1}), this is
\begin{eqnarray}
  \Lambda \left( z_{\ast} \right) & = & \frac{N}{z_{\ast}} f \left(
  \frac{z_{\ast}}{l_1} \right) .  \label{LambdaN}
\end{eqnarray}

This allows us to discuss the perturbativity of the model. If there were no
particles except the UL modes, the scattering of these UL modes would become
non-perturbative at energies of order $4 \pi f$. However, light enough
resonances can tame the amplitudes for scattering of light modes (as a Higgs
boson would), yielding a model that remains perturbative until a higher scale
$\Lambda_{\tmop{IR}}$. It turns out that a resonance spectrum starting at $4
\mathpi f / \sqrt{N_{\tmop{KK}}}$ buys predictive power up to a scale given by
$\Lambda_{\tmop{IR}} \sim 4 \mathpi \sqrt{N_{\tmop{KK}}} f$
{\cite{hep-ph/0111016,Chivukula:2002ej}}. This effective number of KK modes contributing to
a given process is
\begin{eqnarray}
  N_{\tmop{KK}} \left( z_{\ast} \right) & = & Nf \left( \frac{z_{\ast}}{l_1}
  \right) .  \label{NKKz}
\end{eqnarray}
In AdS, because of conformal invariance, this turns out to be constant, and
$N_{\tmop{KK}} = N$. For other warp factors, the effective size of the strong
sector will depend on the energy scale as (\ref{NKKz}). The
$\Lambda_{\tmop{IR}}$ for scattering of light modes corresponds to using
(\ref{LambdaN}-\ref{NKKz}) with $z_{\ast}$ of order {\emdash}but usually
smaller than{\emdash} $l_1$.

The generic picture is then that of Fig.\ref{fig-ng}, where the different
scales are depicted. We have included some numerical values corresponding to
the extreme case of Section \ref{pheno} with the lightest resonance at 600
GeV. Besides the massless photon, the spectrum consists UL modes, to be
identified with the $W^{\pm}$ and $Z$ modes. The KK resonances starts at a
higher scale (of order a few $1 / l_1$), which is parametrically larger than
$M_W$ by a factor $\sqrt{G}$. There are $N_{\tmop{KK}}$ resonances below the
IR cut-off $\Lambda_{\tmop{IR}}$. The resonance spectrum is (approximately)
equally spaced. The whole KK picture would break down at a scale
$\Lambda_{\tmop{UV}}$, which is essentially $\Lambda_{\tmop{IR}}$ times a
blue-shift factor of order $\left( Nl_1 \right) / \left( N_{\tmop{KK}} l_0
\right)$.

\EPSFIGURE[!h]{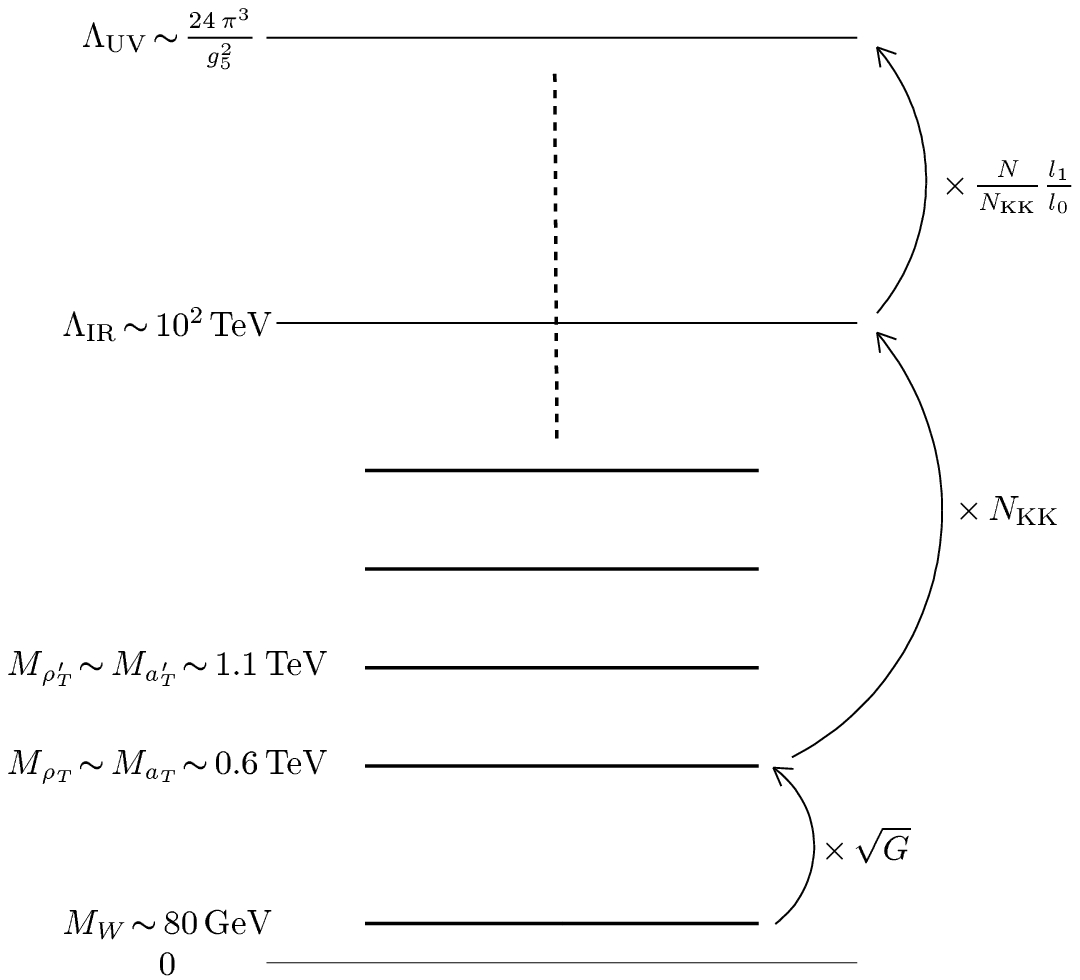}{\label{fig-ng}Schematic representation of the spectrum and relevant
  quantities (not to scale).}

\section{The effect of background fields on $S$}\label{point1}

Here we would like to illustrate claims {\tmem{1.)}} and {\tmem{2.)}} of
Section \ref{our-sol}, namely that there is no prediction for $S$ in 5D models
and that $S$ can change sign. In Section \ref{pheno}, we will show that one
can turn the experimental value of $S_{}$ into predictions for the new physics
sector.

\subsection{Holographic QCD: modifying the axial}\label{modaxial}

Let us jump to GeV physics: we start by the case of Holographic QCD considered by
{\cite{Erlich:2005qh,DaRold:2005ju}}. A bulk field (representing the quark
condensate) triggers chiral symmetry breaking by coupling to the axial sector,
and modifying its profile. Since this distinguishes the vector from the axial
fields, chiral symmetry is broken. See Appendix \ref{rewrite} for details.

What is the effect of an order one change of the condensate (background field
vev) on $S$? The exact derivation is presented in Appendix \ref{appA}, but the
effect is {\tmem{an order one change in the value of S}}. In
Fig.\ref{fig-QCDSoA} we use the model of {\cite{DaRold:2005ju}} which we feed into
our analogue computer using Appendices \ref{rewrite} and \ref{appA}. We plot the value of $S$ as a function of $o_A= \frac{15 \pi^3}{N} \alpha_s \langle \bar{q} q
\rangle^2 l_1^6$.  The specific value $o_A \simeq 16$ used by
{\cite{DaRold:2005ju}} in a fit to QCD data is represented by a star. Note
that the aim of Holographic QCD was not to {\tmem{predict}} the value of $S =
- 16 \pi L_{10}$, but to extract it from data and correlate it with other
observables.

\EPSFIGURE[!h]{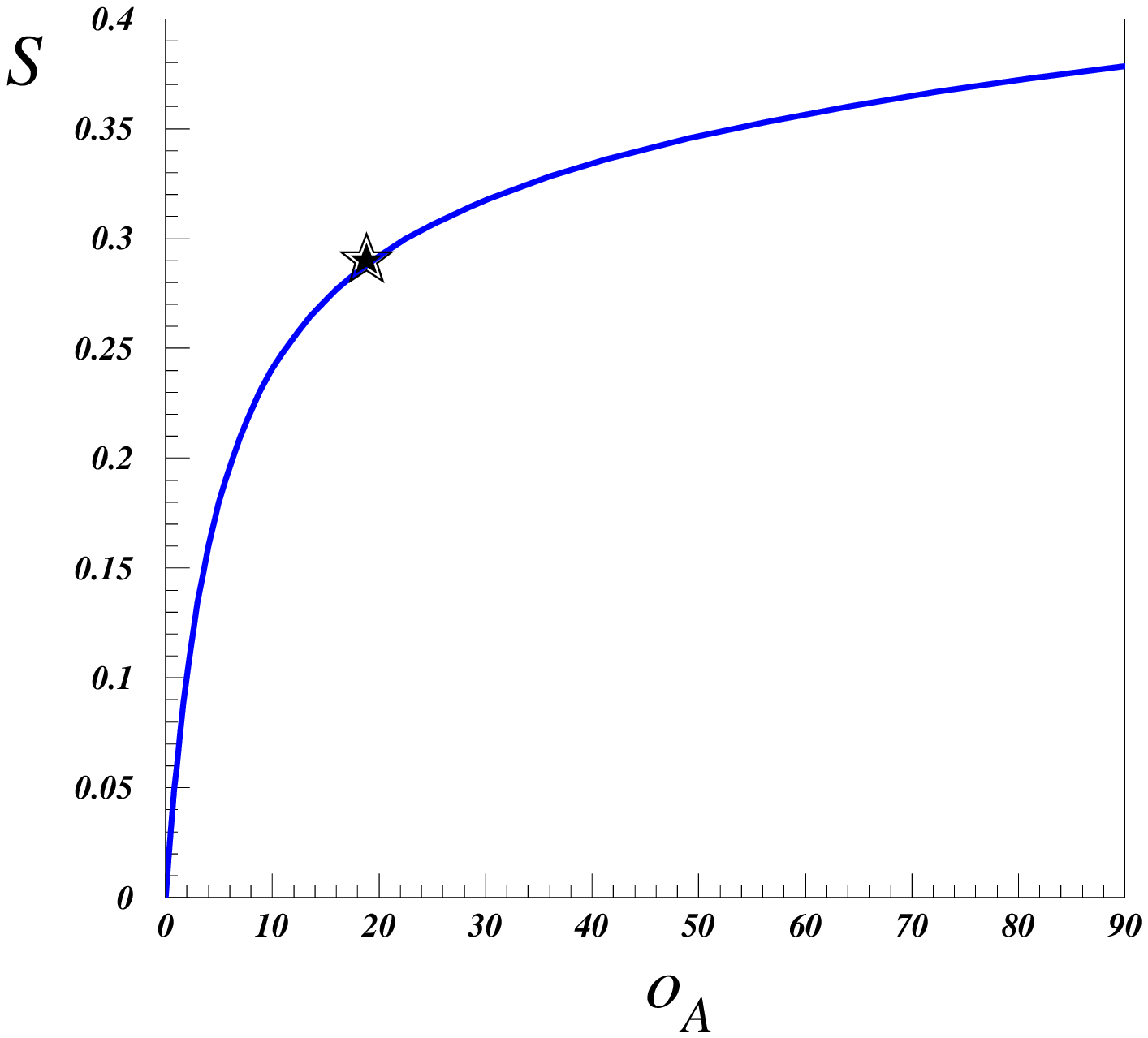,width=10cm}{\label{fig-QCDSoA}$- 16 \pi L_{10}$ for the Holographic QCD model
  of {\cite{daRold:2005zs}}, as a function of $o_A$ ($\xi^2$ in their
  notation). The best fit to QCD in {\cite{daRold:2005zs}} is $o_A \simeq 16$,
  as depicted by the star.}

One thing to notice is that, although one cannot predict the particular
value of $S$, adding the effect of chiral symmetry in the picture
{\tmem{always leads to a positive}} value of $S$. Here we see again how models
with purely rescaled QCD are not able to pass the EWPT unless $N$ is really
small {\emdash}such models would not be computable in the 5D picture
{\cite{Luty:2004ye}}.

The second question one has to address is: how sensitive is the value of $S$ to
the particular modeling of the IR physics? In Fig.\ref{SNoA} we compare the
value of $S$ \ computed in a model with a dynamical scalar coupled to the
axial (as in {\cite{Erlich:2005qh,DaRold:2005ju}}) with the same $S$ computed
by simply adding an exponential profile to the metric itself
(Eq.(\ref{metric2}) with $o_V = 0$). In both cases we have used Dirichlet BC
for the axial fields on the IR brane. $S$ changes by a few percent. This is
also true for the spectrum: changing the parametrization (\ref{metric2}) does
not affect phenomenology much.

\EPSFIGURE[!h]{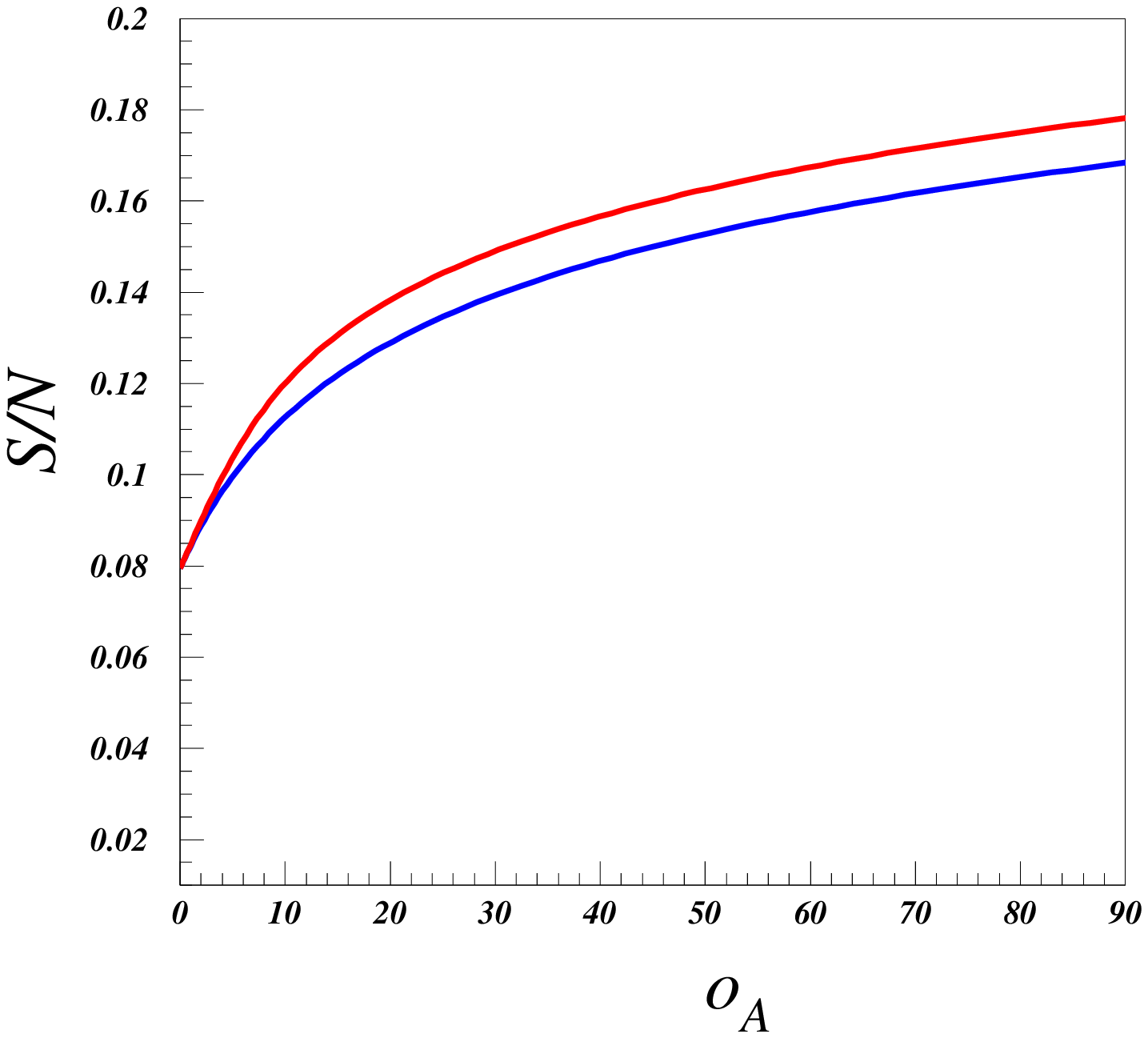,width=10cm}{\label{SNoA}$S / N$ vs $o_A$ for $2 d = 4$. The Figure shows the
  exponential Ansatz (upper curve) and the exact hypergeometric result (lower
  curve). Both cases are computed assuming the large-condensate approximation
  (Dirichlet IR BC).}

\subsection{A simple toy model: modifying the vector }\label{modvector}

Now let us consider a completely different case: imagine it is the vector, not
the axial, who feels the effect of a condensate. We thus set $o_A = 0$ in
Eq.(\ref{metric2}). The value for $S$ can be computed analytically. For
negative $o_V$ we get (see Appendix \ref{appA} and {\cite{Hirn:2006nt}})
\begin{eqnarray}
  S (o_V) & = & \frac{N}{4 \pi}  \left( 1 - \frac{2}{3 d} (\Gamma (0, \nu) +
  \log (\nu) + \gamma_E) \right),  \label{Sov}
\end{eqnarray}
where $\nu = - o_V / \left( 2 d (d - 1) \right) .$

Note that $S$ can be either {\tmem{positive}}, or {\tmem{negative}}. We can
now ask how easy it is to fix $S$ to be very small. The answer depends on the
dimension of the condensate (mass of the 5D field). In Fig.\ref{fig-oVd} we
represent the necessary value of $\left. | o_V \right|$ to yield $S = 0$ from
Eq.(\ref{Sov}). The higher the dimension $d$, the more difficult is for the
field to produce an effect on observables. The natural size for $o$ can be
judged from the Holographic QCD result $o_A \simeq 16$ {\cite{daRold:2005zs}}.
In other words, a high dimension condensate is too peaked towards the IR brane
in order to give a sizable effect. Thinking on localized terms on the IR as
infinite dimension condensates already tells you that they cannot help
lowering the $S$ (their NDA size implies a small effect, unless $N$ is small).

\EPSFIGURE[!h]{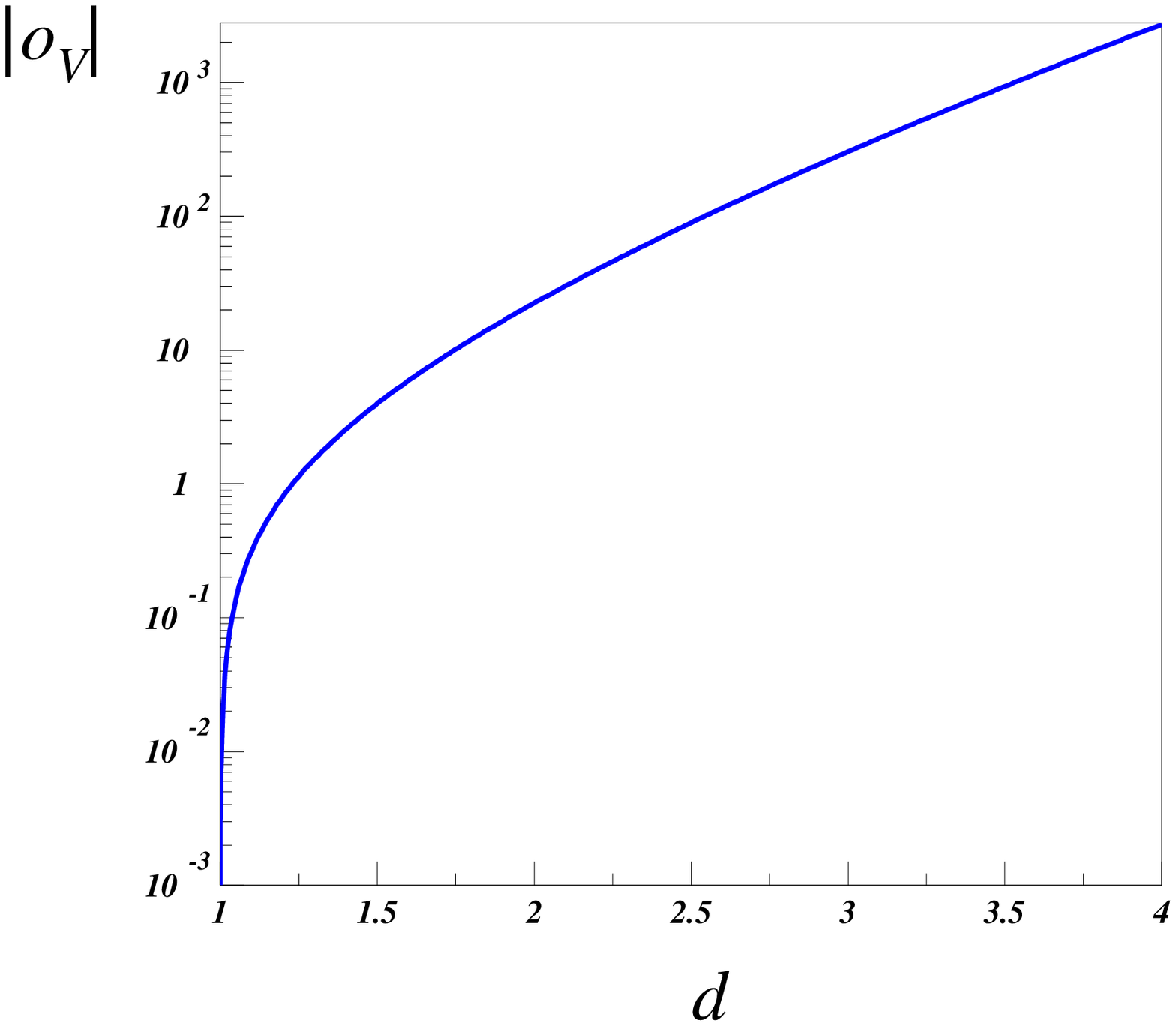,width=10cm}{\label{fig-oVd}Magnitude of $o_V$ necessary to invert the sign of
  $S$, as a function of the dimension $d$, with $o_A = 0$. }

\subsection{A realistic example}\label{realistic}

We saw in Section \ref{modaxial}, Fig.\ref{SNoA} that the value of $S$ is
quite insensitive to the modeling of deviations from conformality. Neither is
the spectrum or the couplings to SM fields. Therefore, we can parametrize the
breaking of AdS conformal invariance as in (\ref{metric2}).

Note that the {\tmem{particular dynamics generating the values of $o_V$ and
$o_A$ is irrelevant for phenomenolgy}} {\footnote{For example, in
{\cite{Hirn:2006nt}} we simply added a $L R$ term parametrizing $w_A - w_V$ at
the quadratic level. }}. We just want to show here an explicit example of a
natural theory leading to these effects.

As an example of dynamics capable of producing such deviations from
conformality, we consider the following action
\begin{eqnarray}
  S & = & \frac{1}{2 \kappa^2} \int d^5 x \sqrt{g}  \left( -\mathcal{R}-
  V_{\phi} (\phi) + \frac{1}{2} g^{M N} \partial_M \phi \partial_N \phi
  \right) \nonumber\\
  & - & \frac{1}{4 g_5^2}  \int \mathd^5 x \sqrt{g} g^{M N} g^{R S} 
  \left\langle L_{M R} L_{N S} + R_{M R} R_{N S} \right\rangle \nonumber\\
  & + & \frac{1}{2 g_5^2}  \int \mathd^5 x \sqrt{g}  \left( g^{M N} 
  \left\langle D_M XD_N X \right\rangle - V_X \left( X \right) \right), 
  \label{Sgrav}
\end{eqnarray}
where $\left\langle \cdots \right\rangle$ means the trace in flavor space, and
$R_{M N} \equiv \partial_M R_N - \partial_N R_M - \mathi [R_M, R_N]$. The
square of the 5D YM coupling $g_5^2$ has dimensions of length. $\kappa$ is the
5D Newton constant related to the curvature $l_0$ and the bulk cosmological
constant $\Lambda$ by $1 / l_0^2 = - \kappa^2 \Lambda / 6$.

Essentially, the action (\ref{Sgrav}) contains the effect of two background
fields: $\phi$ affects gravity $g (x^{\mu}, z)$ and $X$ mostly affects the
Yang-Mills field $A \propto L - R$. In Appendix \ref{appscalar}, we derive
solutions for this model and show that the net effect can be absorbed into
effective metrics. In particular, we show that the effect of $\phi$ and $X$ on
$w_{A, V}$ is the following {\footnote{We neglect the dynamics of the fields
responsible of the modifications {\cite{DaRold:2005vr}}, except the light
modes. See discussion after Eq.(\ref{gensolalpha}).}}:
\begin{enumeratenumeric}
  \item {\tmem{Common background:{\tmem{}}}} the real scalar $\phi$($z$)
  coupled to gravity will produce an effect common to vector and axial ($w_A =
  w_V$). If $\phi$ is non-tachyonic {\footnote{See {\cite{Csaki:2006ji}} for a
  different approach.}}, the effect goes in the direction
  \begin{eqnarray}
    \phi \quad \text{non-tachyonic} & \Longrightarrow & o_V^{\phi} \hspace{1em} =
    \hspace{1em} o_A^{\phi} \hspace{1em} < \hspace{1em} 0,  \label{nontach2}
  \end{eqnarray}
  i.e., it shuts off the IR part of the geometry {\emdash}see Appendix
  \ref{appscalar}.
  
  \item {\tmem{Symmetry-breaking by a bulk scalar:}} charged scalar $X$
  coupled to the axial sector.{\tmem{ At the quadratic level,}} the breaking
  of the $\tmop{SU} (2)_L \times \tmop{SU} (2)_R \longrightarrow \tmop{SU}
  (2)_V$ by a bulk scalar is equivalent to introducing an effective metric for
  the axial channel, and modifying the BCs. The Goldstone bosons eaten by the
  $W, Z$ is a combination of the $A_5$ and of the zero mode of the radial part
  of $X$. This effect predicts a definite sign
  \begin{eqnarray}
    o_V^X \hspace{1em} = \hspace{1em} 0, \hspace{2em} o_A^X \hspace{1em} >
    \hspace{1em} 0 &  & 
  \end{eqnarray}
  and results in $w_A > w_V$.
  
  For example, if the background is AdS and if $X$ has a constant 5D mass,
  $\langle X \rangle$ is a power-law
  \begin{eqnarray}
    \langle X \rangle & = & \sigma z^d, 
  \end{eqnarray}
  and we get
  \begin{eqnarray}
    w_X (z) & = & \frac{l_0}{z} _0 F_1 \left( ; \frac{d - 1}{d} ;
    \frac{\sigma^2 l_0^2 z^{2 d}}{2 d^2} \right)^2  \hspace{1em} \underset{d =
    2}{=} \hspace{1em} \frac{l_0}{z} \tmop{Cosh}^2 \left( \frac{\sigma z^2}{2}
    \right) . 
  \end{eqnarray}
  \item Adding several fields of scaling dimensions $2, 3, 4 \ldots d$ would
  have an effect on the metric suppressed by $(z / l_1)^{2 d}$. The lower the
  dimension, the more the deviation from AdS extends inside the bulk. The
  effect is of course maximum on the IR brane, but still there it is
  suppressed by $d (d - 1)$ as the dimension of the condensate increases. See
  Appendix \ref{NDA}.
\end{enumeratenumeric}
To conclude, {\tmem{1.) }}for phenomenological purposes, one only needs to
consider the effect of the lower dimension condensates, {\tmem{2.)}} the
vector channel is only affected by the neutral scalar $\phi$, with a definite
sign
\begin{eqnarray}
  o_V & < & 0,  \label{oVneg}
\end{eqnarray}
whereas the axial channel is also affected by the charged scalar, and the two
effects compete, resulting in
\begin{eqnarray}
  o_A & > & o_V .  \label{oAbigger}
\end{eqnarray}
This is also Witten's positivity condition obtained from the spectral
functions of 4D theories {\cite{Witten:1983ut}}.

\section{Parameter scan for $S$}\label{results}

In this section we study the parameter space that leads to small $S$. The key
point is that one can study this issue and correlate it with the spectrum.
Thanks to our analogue computer, this can be done without going into the
detailed dynamics that produced the deviations. Studying these correlations
will be the point of Section \ref{pheno}.

The importance of encoding the effects of various background fields into
effective metrics is that it simplifies the task of computing (4D)
observables. Many 4D quantities involve contributions from{\tmem{ all}} KK
modes. Using the effective metrics, such sums can be expressed as simple
integrals over the fifth dimension. To summarize, a Sum Rule (SR) works as
follows
\begin{eqnarray*}
  \text{relevant 4D quantities} & = & \sum_{\tmop{KKs}} \bignone \text{KK
  properties} \\
  & = &  \text{geometrical factor} .
\end{eqnarray*}
The beauty of the SR is to relate the sum over KK contributions with a pure
geometrical factor that can be computed with just the knowledge of the metric.
This is an advantage because sampling KK properties over a whole parameter
space would be a herculean task. Namely, if deviations of AdS are included,
one would have to solve numerically the equations of motions for at least the
low-lying states and extract the masses and couplings {\footnote{In addition,
it will turn out that, for models yielding $S \simeq 0$, one needs to take
many states into account before noticing that the vector and axial
contributions to $S$ cancel out.}}.

The $S$ parameter is a good example of a SR: we have
\begin{eqnarray}
  S & = & 4 \pi \sum_n \bignone f_{V_n}^2 - f_{A_n}^2 \nonumber\\
  & = & \frac{N}{3 \pi}  \int_{l_0}^{l_1}  \frac{dz}{l_0} (w_V (z) - w_A (z)
  \alpha^2 (z)),  \label{SSR}
\end{eqnarray}
where $\alpha (z)$ is the wavefunction of the GBs and it is again purely
geometrical
\begin{eqnarray}
  \alpha (z) & = & 1 - \frac{\int^z_{l_0} \frac{dz'}{w_A
  (z')}}{\int_{l_0}^{l_1} \frac{dz}{w_A (z)}},  \label{alpha}
\end{eqnarray}
and $0 < \alpha < 1$. If instead of breaking chiral symmetry by BCs, one uses
the bulk scalar $X (z)$, $\alpha (z)$ is slightly modified (see Appendix
\ref{appA}, Eq.(\ref{gensolalpha})).

The result (\ref{SSR}) can also be understood by using the original
definition for $S$ {\cite{Peskin:1991sw}}
\begin{eqnarray}
  S & = & \left. 2 \pi \frac{d}{d Q^2} \left( Q^2 \Pi_V \left( Q^2 \right) -
  Q^2 \Pi_A \left( Q^2 \right) \right) \right|_{Q^2 = 0}, 
\end{eqnarray}
which is the difference between the kinetic terms generated for the $V$ and
$A$ sources. These two terms correspond to the two terms in (\ref{SSR}) as can
be understood from the following. Non-zero sources generate a field $\Phi_{V,
A} \left( z \right)$ in the bulk, yielding a 4D kinetic term $\int \mathd z /
g_5^2 \bignone w_X \left( z \right) \Phi_X \left( z \right)^2$. Now, the
$\Phi$'s obey the standard IR BCs, but are subject to the UV normalization
appropriate for sources $\Phi_{V, A} \left( z \right) = 1$. Solving for the
massless wave equation, one finds $\Phi_{V, A} \left( z \right) = 1, \alpha
\left( z \right)$ respectively, which leads to the previous result
(\ref{SSR}).

Note that $S$ is insensitive to the UV cutoff $l_0$. This is what one should
expect for a low-energy quantity coming from the strong sector {\footnote{See
{\cite{Piai:2006hy}} for a different approach (by keeping $l_1 / l_0 \sim
6$).}}.

\EPSFIGURE[!h]{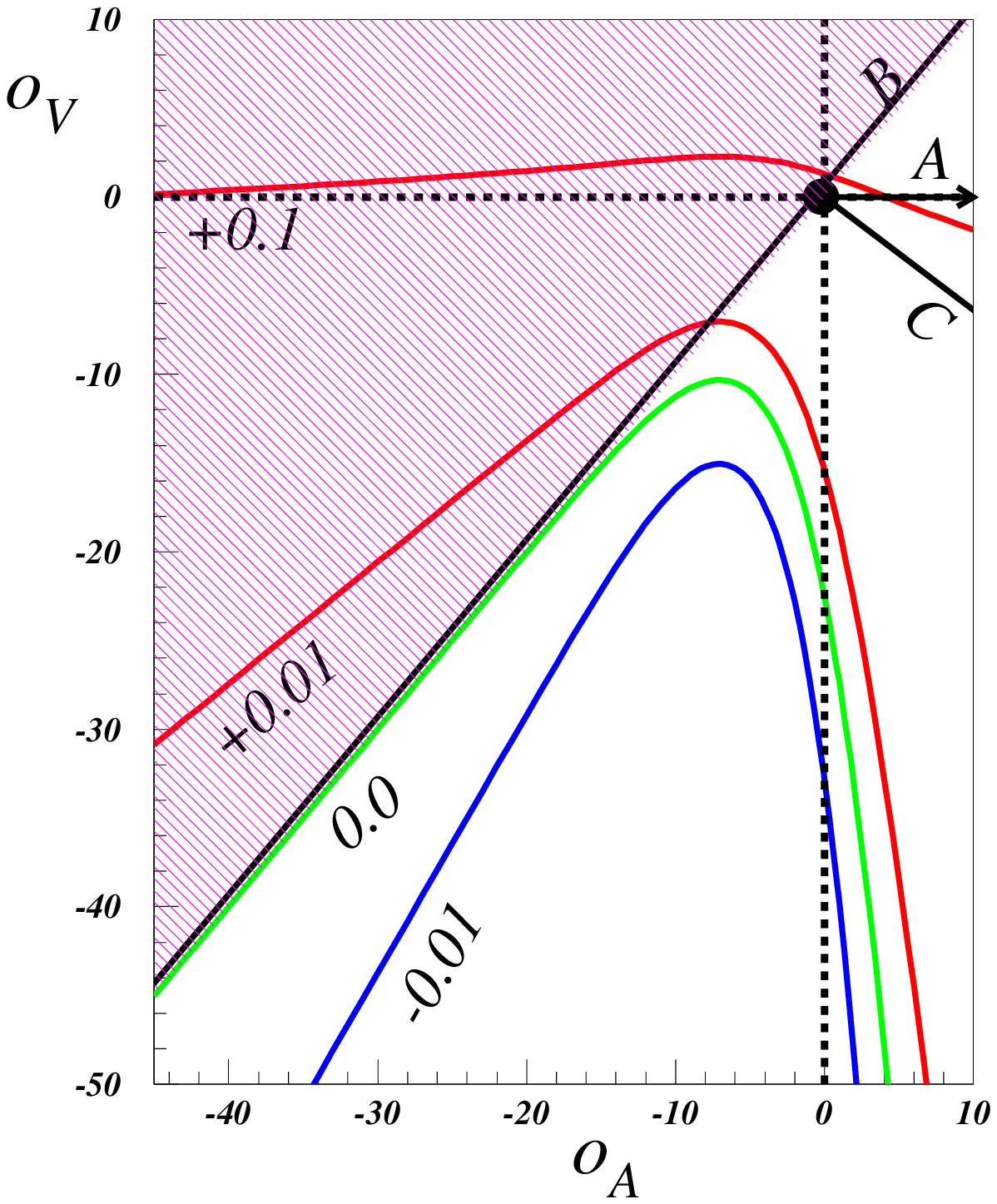,width=10cm}{\label{fig-S/Ncontour}Contour lines for $S / N$ in the $\left( o_A,
  o_V \right)$ plane, for the case $2 d = 4$. The dot at the origin represents
  the original warped Higgsless model {\cite{Csaki:2003zu}}. Line B
  corresponds to having $w_A = w_V$, but with \ warping different from AdS, as
  in {\cite{Barbieri:2003pr}}. Line C denotes the class of models respecting
  the QCD factorization relation $o_V = - 7 / 11 o_A < 0$. The arrow A
  represents models with condensates only in the axial channel
  {\cite{daRold:2005zs}}. The shaded region is forbidden by Witten's
  positivity condition.}

In our explicit example, the $\phi$ field affects both axial and vector while
the effect of $X$ goes in the opposite direction leading to the conditions
$o_V < 0$ and $o_A > o_V$ (\ref{oVneg}-\ref{oAbigger}). We are fortunate that
the region of $S \leqslant 0$ lies in that region, see
Fig.\ref{fig-S/Ncontour}. In that Figure, we show the region in parameter space where $S$
changes sign. In pure AdS, $S = N / 4 \pi$ {\cite{Csaki:2002gy}}. The authors
of {\cite{Barbieri:2003pr}} realized that increasing the (common) warping
would not change the sign of $S$, as you can realize by looking at the
expression of $S$, Eq.(\ref{SSR}), with $w_A = w_V$ and noting that $\alpha
\leqslant 1$. These authors also noted that one can make $S$ small by going to
the lower-left part of the diagonal (line B) in Fig.\ref{fig-S/Ncontour}, but
that this would require a low $N$.

Another direction explored by the authors of {\cite{daRold:2005zs}} is to
increase $o_A$: this is depicted in Fig.\ref{fig-S/Ncontour} by the arrow A
pointing along the $x > 0$ axis. The model of {\cite{Piai:2006hy}} should also
lie on that arrow, but it treats the UV differently.

\section{Phenomenology}\label{pheno}

From the study of the $\left( o_A, o_V \right)$ parameter space performed in
Section \ref{results}, we found a region corresponding to $S \simeq 0$. We now
extract the characteristics of models in that region.

To go further, we focus on the extreme case with the lightest possible
resonances, $M_{a_T} = 600 \tmop{GeV}$. Having such a light KK compared to
$M_W$ improves perturbativity (increases $N_{\tmop{KK}}$) as depicted in
Fig.\ref{fig-ng}. We also have to match the Fermi constant $G_F$, or
equivalently $f = 246 \tmop{GeV}$. This allows us to fix enough parameters to
draw the exclusion plot for $- 0.5 < S_{\tmop{tree}} < 0.1$ in
Fig.\ref{fig-Sband}.

\EPSFIGURE[!h]{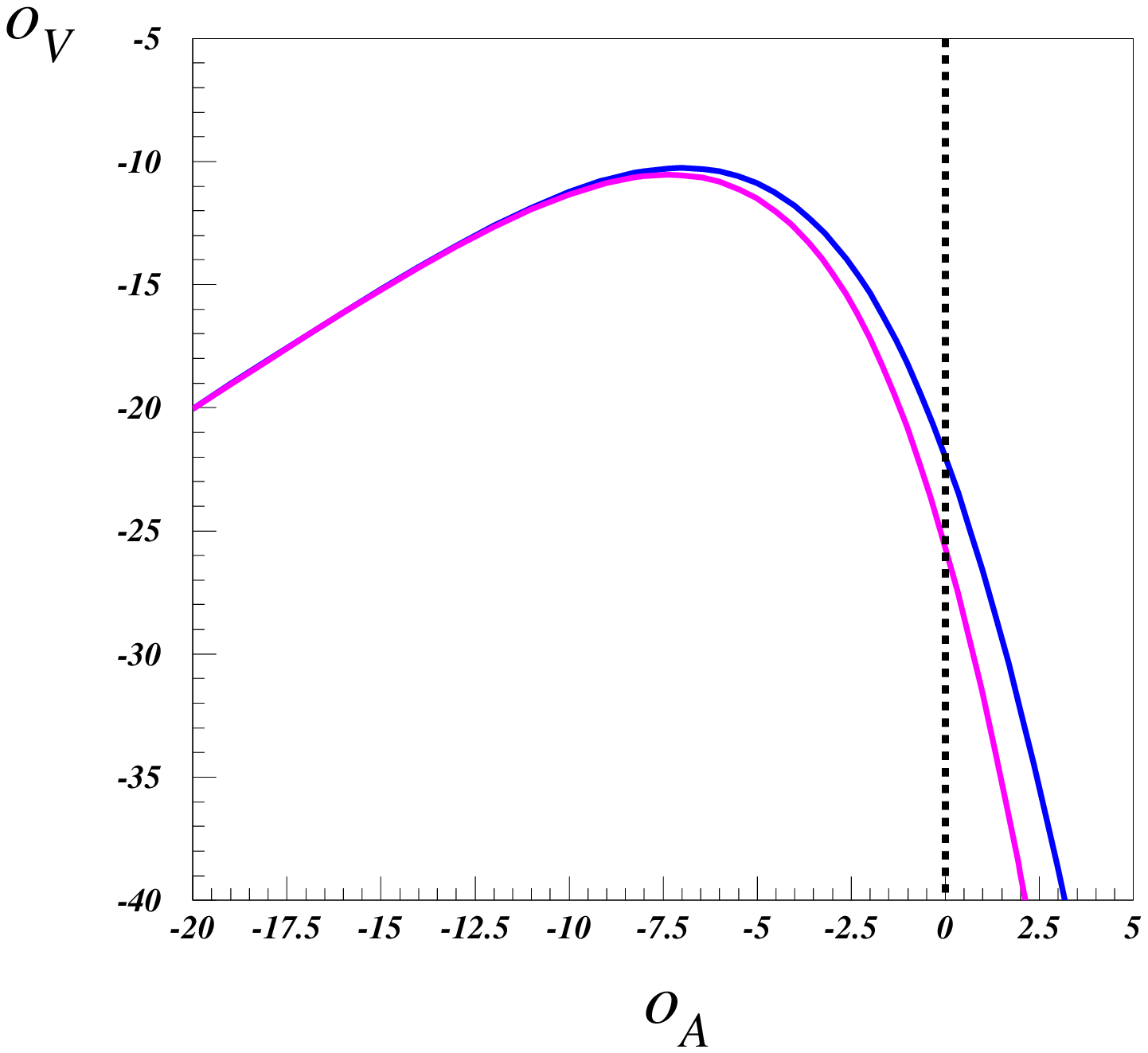,width=10cm}{\label{fig-Sband}The allowed parameter space in the $\left( o_A,
  o_V \right)$ plane in order to satisfy $- 0.5 < S < 0.1$, after imposing
  $M_{a_T} = 600 \tmop{GeV}$.}

In the (narrow) left part of the band, one ends up in a
situation where vector and axial resonances are nearly degenerate. In the
right part of the band ($o_A > - 10$), the spectrum is inverted respect to the
QCD case: the axial resonance is lighter than the vector one. This is depicted
in Fig.\ref{fig-massratios}, (see \ where we plot the ratio $M_{\rho} /
M_{a_T}$ as a function of $o_A$, along the line of $S = 0$.

\EPSFIGURE[!h]{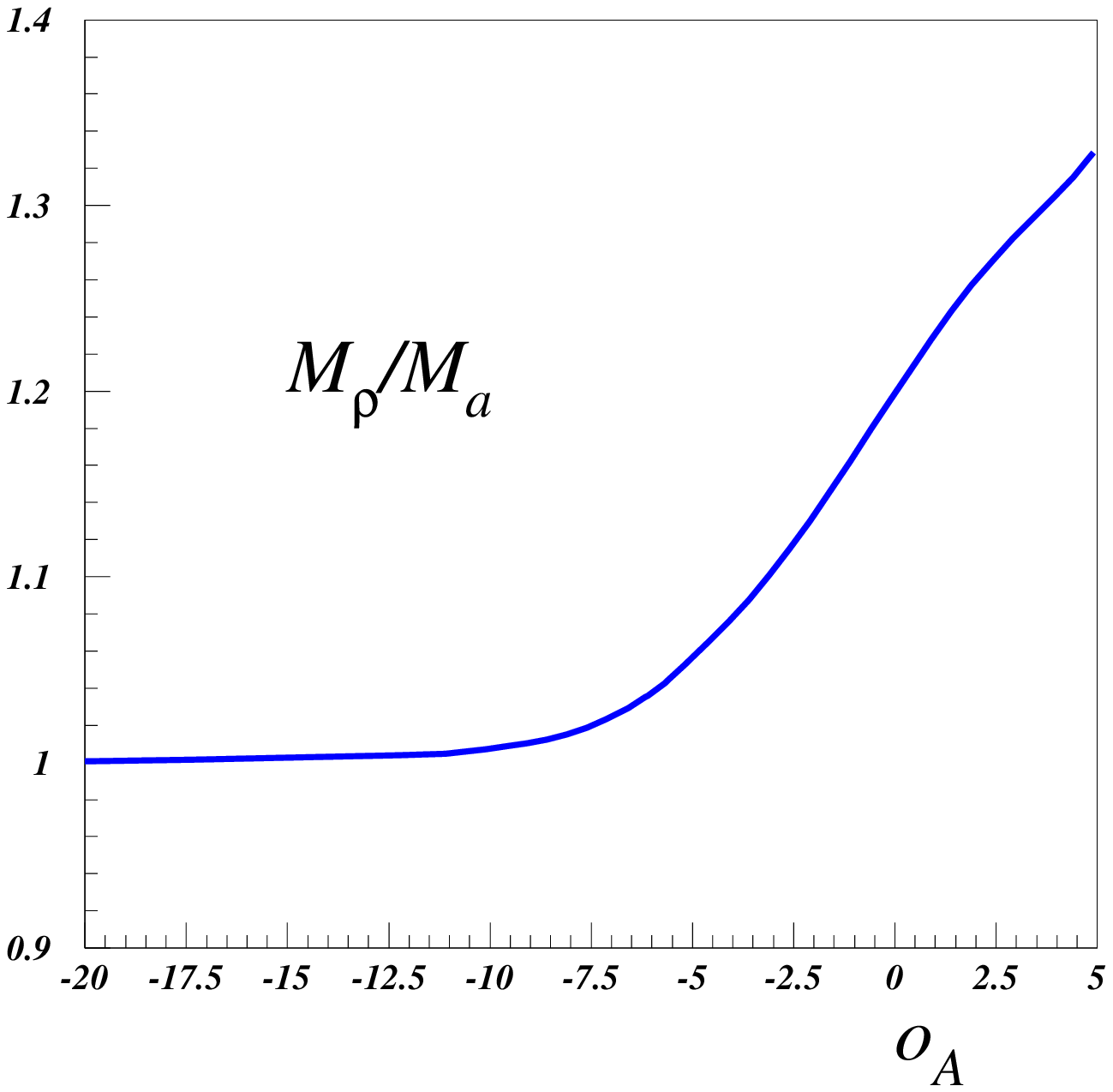,width=9cm}{\label{fig-massratios}Plot of the ratio $M_{\rho_T} / M_{a_T}$ along the line of $S = 0$, as a function of $o_A .$}

 In Fig.\ref{fig-spectrum} we depict the two situations we have just explained:
degenerate or inverted spectrum for $o_A \lessgtr - 10$.

\EPSFIGURE[!h]{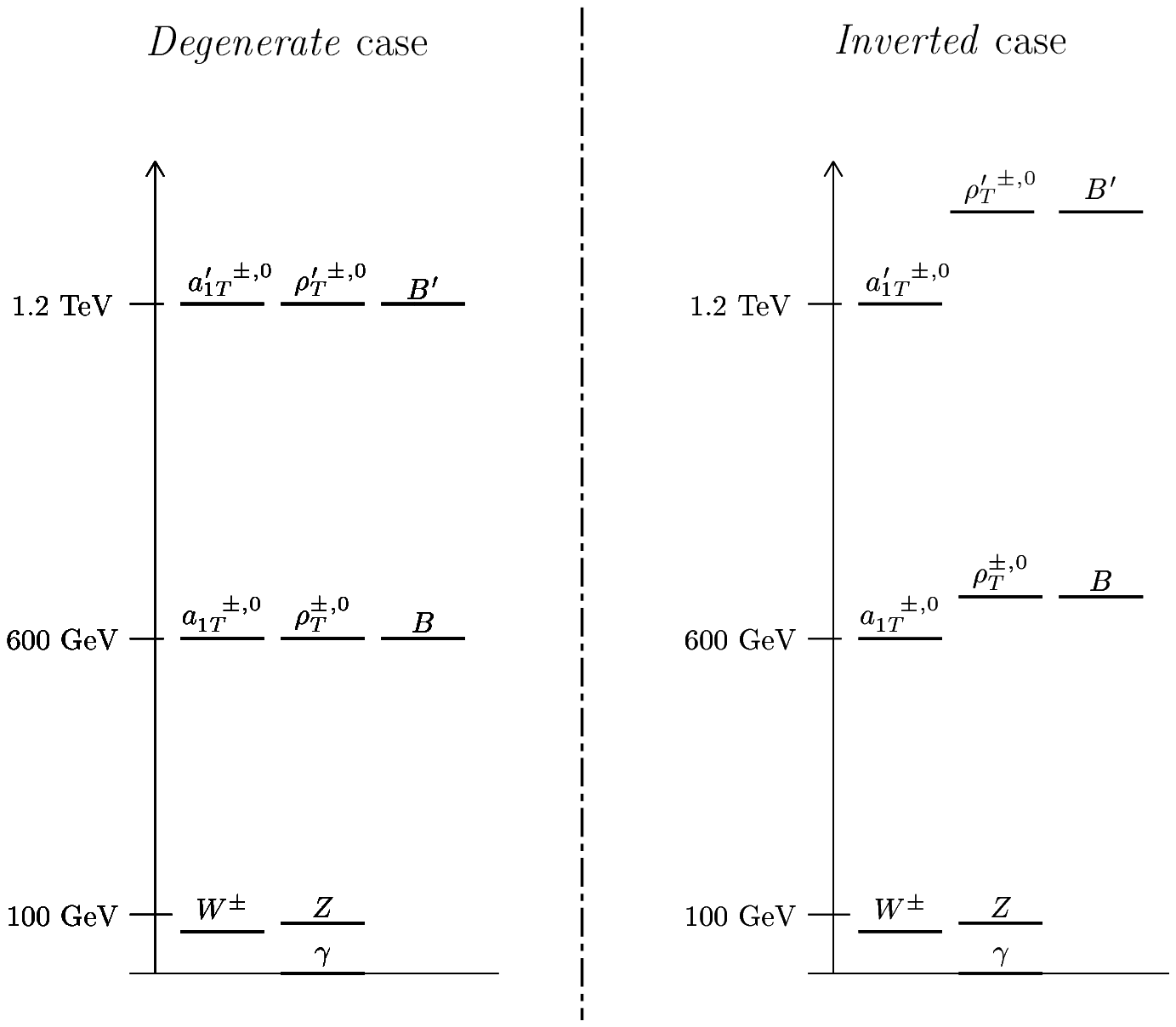,width=9cm}{\label{fig-spectrum}Schematic depiction of the spectrum (not to
  scale). The $B$ resonances are the excitations of the $\mathrm{U}(1)_{B-L}$ field.}

Note that, imposing $S = 0$ would imply an exact relation between $o_V$ and
$o_A$, $o_V = f \left( o_A \right)$. Passing the experimental constraint
requires this relation to be fulfilled only approximately, as shown in
Fig.\ref{fig-Sband}. Obviously, an improvement in experimental constraints
would select a narrower region of parameter space, but let us anyway quantify
the adjustment. We see that, for the inverted case $o_A \gtrsim - 10$, $o_V$
only needs to be equal to $f \left( o_A \right)$ within 10\%. The adjustment
between the two potentials $V_X$ and $V_{\phi}$ to obtain this relation is at
the same level as the that of the radion potential
{\cite{Goldberger:1999uk,Goldberger:1999un}}. For $o_A \lesssim - 10$, $o_V$
would have to be equal to $f \left( o_A \right)$ within 1\%. Remember however,
that this is the case where vector and axial resonances are degenerate, which
could come in 4D from a symmetry {\cite{hep-ph/9502247,Appelquist:1999dq,hep-ph/0405188}}.

We have focused up to now on $S$, which involves contributions from all
resonances, and could thus be expressed through a SR. The same is true for
$G_F$. All of this is done automatically within the analogue computer. Since
{\tmem{the effect of background fields can be encoded into a effective metric,
at the quadratic level, we can also use this parametrization to extract the
decay constants and the masses.}} Table \ref{table} shows the successive steps
that lead to predictions. In that case, we have to proceed again for each
point in parameter space.

\begin{table}[!h]
  \begin{tabular}{ccccccc}
    Step &  & Requirement &  & Parameter to set &  & Predictions\\
    &  &  &  &  &  & \\
    0 &  &  &  & Choose $o_A = 0$ &  & \\
    &  &  &  &  &  & \\
    1 &  & $M_{a_T} = 600 \tmop{GeV}$ &  & $l_1 \simeq 6.4 \tmop{TeV}^{- 1}$ &
    & $M_{A_{2, 3, 4}} \simeq 1.1, 1.6, 2.1 \tmop{TeV}$\\
    &  &  &  &  &  & \\
    2 &  & $S = 0$ &  & $o_V \simeq - 22.5$ &  & $M_{V_{1, 2, 3, 4}} \simeq
    0.7, 1.35, 1.9, 2.4 \tmop{TeV}$\\
    &  &  &  &  &  & \\
    3 &  & $f = 246 \tmop{GeV}$ &  & $N = 146$ &  & $f_{V_{1, 2, 3, 4}} \simeq
    13.8, 8.7, 1.9, 2.4$\\
    &  &  &  &  &  & $f_{A_{1, 2, 3, 4}} \simeq 12.3, 9.0, 7.5, 6.5$\\
    &  &  &  &  &  & \\
    4 &  & $M_W = 80.4 \tmop{GeV}$ &  & $\log \left( l_1 / l_0 \right) \simeq
    4.5$ &  & resonance isospin splittings, $T$
  \end{tabular}
  \caption{\label{table}Step-by-step flowchart for our particular benchmark
  model with $S = 0$.}
\end{table}

The 5D parameters necessary to describe the new physics sector are the
coupling constant $1 / N$, the scale of KK resonances $1 / l_1$, and the two
condensates $o_{A, V}$
\begin{eqnarray*}
  5 D \tmop{parameters} : N, l_1, o_A \tmop{and} o_V . &  & 
\end{eqnarray*}
The procedure we follow to fix them in terms of 4D parameters is to use the
value of $f$ and fix the lowest resonance to be at $600 \tmop{GeV}$ (bounds
from TeVatron {\cite{Abe:1997fd}} and LEP {\cite{Barbieri:2004qk}}). Thus, the
value of $N$ and $l_1$ are just functions of $o_A$
\begin{eqnarray*}
  f = 246 \tmop{GeV}, M_{W'} \sim 600 \tmop{GeV} \Longrightarrow N (o_A), l_1
  (o_A) . &  & 
\end{eqnarray*}
We also set $S$ within the experimental range, leading to the determination of
$o_V$ as a function of $o_A$,
\begin{eqnarray*}
  S & \Longrightarrow & o_V (o_A) .
\end{eqnarray*}
 Note that in order to obtain a good approximation
for $S$ by summing up resonances, one needs to take into account the
resonances up to  $\mathcal{O}(10)$  TeV.

This predicts any other observable in terms of one parameter, $o_A$. On the
other hand, a natural potential roughly sets $|o_A | \lesssim 100$. To
illustrate how this proceeds, we show in Table \ref{table} the various steps
for a model with $S = 0$. To fix numbers, we need to pick one value for $o_A$.
We choose $o_A = 0$ for simplicity. Note that, except for step 4, we do not
need to specify $l_0$: everything is finite in the limit $l_0 \longrightarrow
0$, and would only receive small corrections. It is the value of $M_W$ that
sets $l_0 / l_1$.

\section{4D interpretation}\label{4D}

\subsection{UV independence and IR robustness}\label{robust}

The $S$ parameter can be written as the difference between a vector and an
axial contribution, see (\ref{SSR}). While both terms in (\ref{SSR}) are
dominated by the UV, the difference is finite as $l_0 \longrightarrow 0$, as
should be: the $S$ parameter is insensitive to the $\tmop{UV}$ details of the
model, since the chiral symmetry is restored at high energies. This is
embodied in the 5D model by
\begin{eqnarray}
  w_V \left( l_0 \right) & = & w_A \left( l_0 \right), 
\end{eqnarray}
so that the UV does not contribute to the $S$ parameter, as can be seen in
Fig.\ref{fig-Sz}. 

\EPSFIGURE[h]{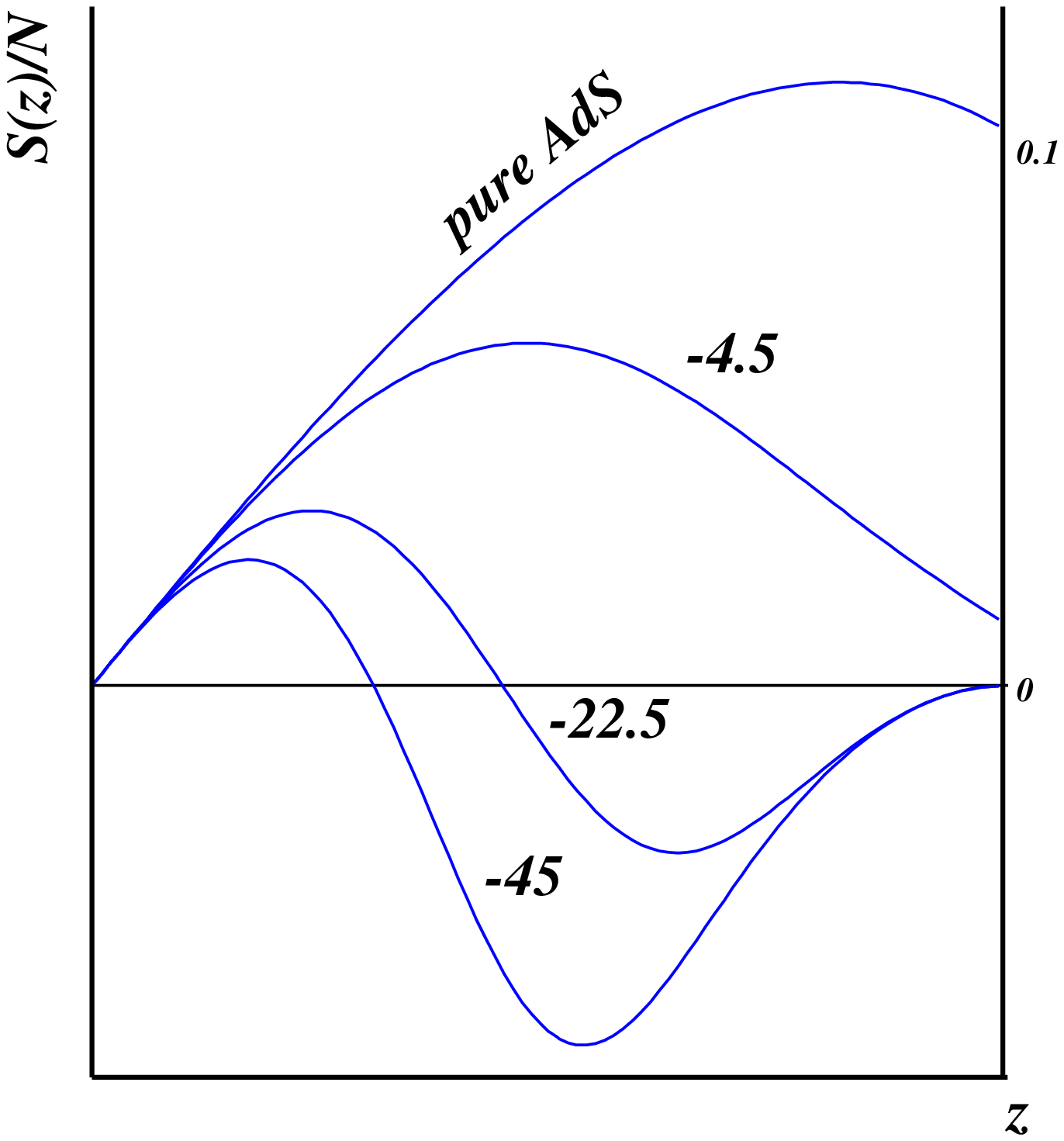,width=10cm}{\label{fig-Sz}Value of the integrand yielding $S$ in the sum rule
  (\ref{SSR}). Indicated is the value of $o_V$, assuming $o_A = 0$ for
  simplicity.}

In this Figure, we show the value of $w_V - w_A \alpha^2$ as
a function of the bulk coordinate $z$. Notice how, in the AdS case, the
contributions come mostly from the IR region, whereas for the cases of
interest, (smaller) contributions compete against each other, and come from
the whole bulk. As a consequence, the low-energy quantity $S$ is independent of
the behavior of the two-point functions at very high energies. This also
implies that we will not be sensitive to the high resonances, and to whether
their spectrum follows the Regge behavior or not. In summary, the 5D model
needs to match the assumed OPE of the two-point functions only for
intermediate energies, not for asymptotically large ones.

Another key point is that the precise form of the deviations near the IR is
not important. Indeed, the integral expression (\ref{SSR}) for $S$ receives
most of its contributions from the bulk. Therefore, a strong suppression of
the metric near the IR is not important for the result. What is essential is
that the condensate in the vector channel be large enough for $z \sim l_1 /
\tmop{few}$. We have indeed checked the robustness of our results when using
different Ans\"atze for the metric instead of (\ref{metric2}) (see for example
Fig.\ref{SNoA}) {\footnote{In this sense, the discussions of
{\cite{Csaki:2006ji}}, putting the emphasis on the extrapolation of the OPE to
the IR in order to generate confinement, and {\cite{Karch:2006pv}}, describing
the dependence of the spectrum of heavy resonances on the shape of the IR
cut-off are not our concern when discussing the $S$ parameter.}}.

Finally, note the following from Fig.\ref{fig-Sz}. In the AdS case, the
contributions to $S$ come from the IR. In that case, $S$ can be estimated by
including only the contribution from the lightest resonance. For $o_V \simeq -
22.5$, which leads to $S \simeq 0$, all the intermediate energies contribute
{\emdash}and cancel out. In this sense, the result for $S$ includes more
contributions from intermediate energies than in the QCD case: one needs to
sum up resonance contributions up to $\mathcal{O}(10)$ TeV to realize that $S$ cancels out. This is
somewhat expected from the 4D side
{\cite{Lane:1993wz,Appelquist:1998xf}}. 

\subsection{Purely 4D argument}\label{Witten}

In this Section, we describe in which way the condensates and the $S$
parameter are correlated, by simply considering the left-right two-point
function. This explains the result of Section \ref{results}, independently of
the 5D modeling. \ This can be done with the help of Fig.\ref{fig-PiLR}, where
$Q^2 \Pi_{L R} \left( Q^2 \right)$ is depicted: the asymptotic behavior of the
curve is given by
\begin{eqnarray}
  Q^2 \Pi_{L R} \left( Q^2 \right) \underset{Q^2 \longrightarrow + \infty}{=}
  \frac{\left\langle \mathcal{O}_V -\mathcal{O}_A \right\rangle}{2 Q^{2 \left(
  d - 1 \right)}} & < & 0,  \label{asymp}
\end{eqnarray}
which has to be negative in order to fulfill Witten's positivity condition,
which states that the whole function should be negative
{\cite{Witten:1983ut}}. Also, the GB decay constant can be defined from the
intercept at the origin
\begin{eqnarray}
  \left. Q^2 \Pi_{L R} \left( Q^2 \right) \right|_{Q^2 = 0} & = & - f^2 . 
\end{eqnarray}
\EPSFIGURE[h]{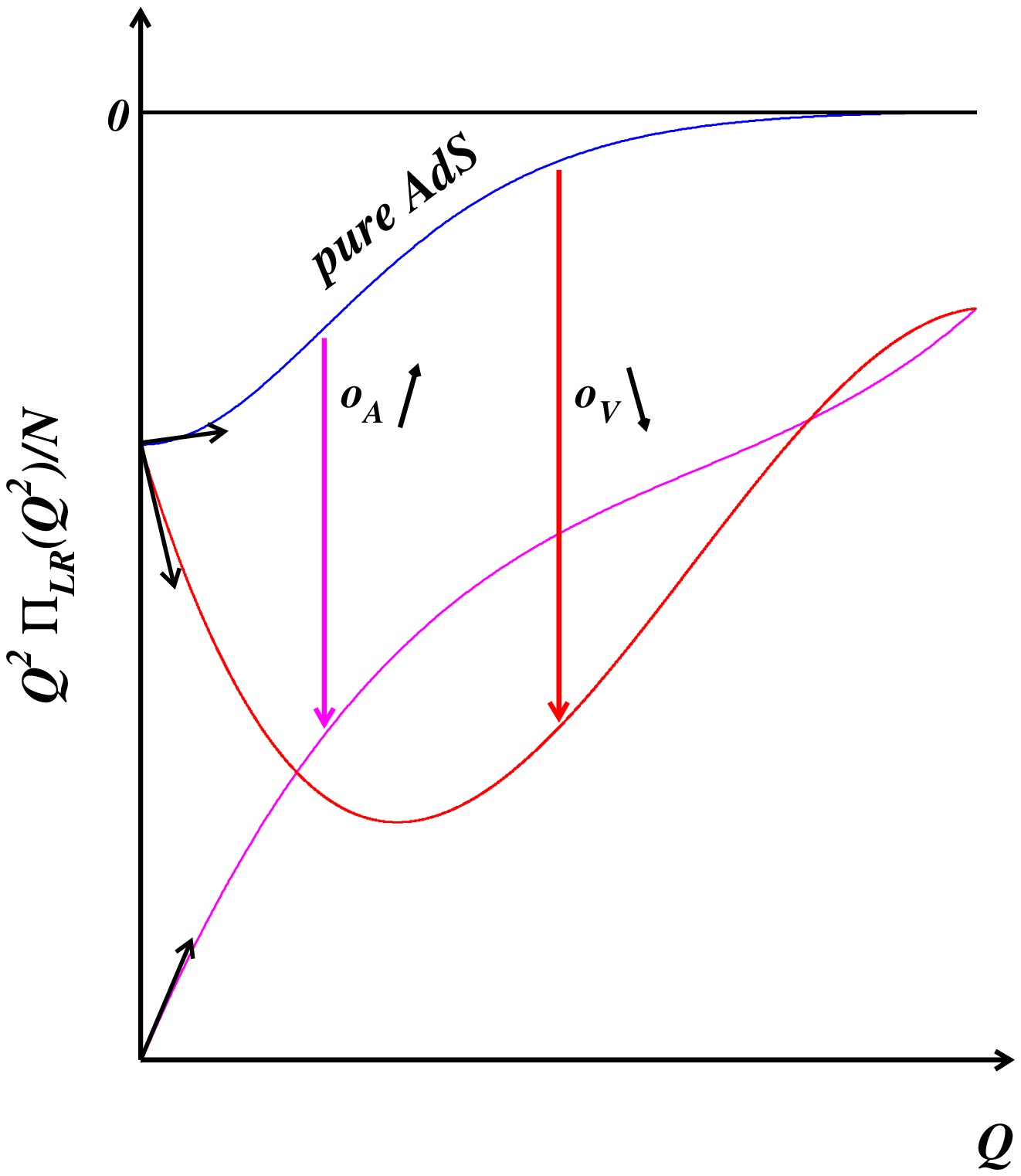,width=9cm}{\label{fig-PiLR}Cartoon of the left-right two-point function,
  showing the influence of axial and vector condensates on the slope at the
  origin, $S$.}

Like any 4D model, the present model provides an interpolant between these two
regimes, using a set of spin-1 resonances, while satisfying basic
field-theoretical requirements. Also, as a bonus compared to earlier models,
it includes the perturbative behavior of both two-point functions $\Pi_V$ and
$\Pi_A$ {\tmem{separately}}. This has an incidence, as the result does
{\tmem{not}} depend only on the difference $o_V - o_A$, but on the two
variables $o_V$ {\tmem{and}} $o_A$ {\footnote{To get an idea of the way things
work in simple 4D models see Appendix \ref{res-sat}.}}.

Fig.\ref{fig-S/Ncontour} shows that obtaining $S < 0$ requires $o_V < 0$. This
can be understood in general, without resorting to the 5D model as follows.
Start from the fact that the standard QCD case, or even the case $o_V = o_A =
0$ lead to $S > 0$. The upper curve in Fig.\ref{fig-PiLR} represents $Q^2
\Pi_{L R} \left( Q^2 \right)$ for this latter case. Making $o_A$ positive
would bring the curve down in the IR according to the asymptotic behavior
(\ref{asymp}). This is however not enough to make the slope at the origin
($S$) negative. This is because the intercept $- f^2$ also goes down with
larger $o_A$: $f^2$ is the decay constant of the would-be GBs, and is
therefore sensitive to the condensates in the axial channel. The outcome is
depicted by the second curve in Fig.\ref{fig-PiLR}, which also has a positive
slope at the origin. On the other hand, decreasing $o_V$ would also bring the
curve down in the UV, but this time without modifying the intercept at the
origin. In this case, one can get a negative slope at the origin, $S < 0$,
third curve in Fig.\ref{fig-PiLR}. The same kind of reasoning would also show
that the effect is greater with a low-dimension condensate.

\section{Conclusions}\label{conclusions}

We have presented an effective parametrization of quadratic interactions
between spin-1 KK resonances, in a 5D model defined on an interval. The key
point is that, at the quadratic level for the gauge fields, any coupling of a
background field can be recast as an effective metric. This is even true in
the case of background fields with light excitations
{\cite{Erlich:2005qh,daRold:2005zs}}. We have also displayed how the rewriting
works in examples of explicit 5D models including background fields.

Though this rewriting may be performed for one's favorite model, we focus on
the next step: we start from our generic parametrization, and consider the
physical consequences independently of the details of the underlying dynamics.
In this respect, the parametrization used here is an {\tmem{analogue
computer}}: it allows us to study the correlations between observables. Here,
we considered the interplay between the spectrum and the $S$ parameter
{\footnote{In a paper in preparation we explore other correlations between
observables and the $S$.}}. We have performed this analysis with the simplest
modeling of the IR cut-off, and explored the parameter space that leads to a
phenomenologically viable $S \simeq 0$. It turns out that the results do not
depend on the deep IR modeling, but rather on the behavior for the whole range
of intermediate scales. This is expected from discussions of walking in 4D
technicolor {\cite{Lane:1993wz,Sundrum:1991rf,Appelquist:1998xf}}. The result
for the low-energy parameter $S$ is also UV-insensitive, as should be.

There is a common lore that the $S$ parameter constraint excludes strong
interactions as the source of EWSB. Our results strengthen the objections to
this claim. What is indeed true is that strong interactions do generate large
{\emdash}proportional to $N${\emdash} contributions to $S$. Still, these
contributions do not have a fixed value, but strongly vary in the parameter
space we have explored. In fact, we find that cancellations between the vector
and axial contributions to $S$ do occur for reasonable values of these input
parameters.

Let us address the question of fine-tuning. The analogue computer is a tool
that parametrizes the effects of background fields on phenomenology without
relying on particular dynamics. In this context, questions on fine-tuning of
parameters are meaningless: whether there is a model that predicts some
{\tmem{particular}} values for the metric is out of the scope of this
approach. The point of a 5D model is not to {\tmem{predict}} the value of $S$,
but to {\tmem{correlate}} the experimental value for $S$ with properties of
the strongly coupled sector observables. We have examined the region of
parameter space corresponding to $S \simeq 0$, and shown that it corresponds
to having the axial resonances either degenerate with (as already considered
in {\cite{Appelquist:1999dq}}), or lighter than the vector ones. We stress
that the scenarios we have considered have a sufficiently large $N$ to enjoy a
weakly-coupled mesonic/5D description until a few tens of TeV. To achieve
this, the lightest resonance (the techni-$a_1$) should appear below a TeV
(maybe as low as 600 GeV), which is now allowed since the constraint on $S$ is
lifted. Also, we repeat that the mechanism of cancellation can be equally well
applied to the present extreme Higgsless case as to composite or gaugephobic
Higgs models.

Another point of our analogue computer is that its effective parameters are
directly related to terms in the 4D OPE. This allows us to point to specific
directions in the space of 4D strong interactions. Admittedly, devising a 4D
mechanism that generates these effective parameters dynamically will be a much
harder enterprise. Still the following statements can be made. To obtain $S
\lesssim 0$, one needs a significant departure from AdS in the bulk
{\emdash}{\tmem{not just on the IR brane}}. In the 4D picture, this is
tantamount to having a low-dimension condensate with a sizable magnitude. It
may be that walking effectively produces such a low-dimension scaling, via the
large anomalous dimension that the quark condensate acquires. To clarify a
possible connection with walking, we have thus shown how to translate the
various scales of the 5D model into those of extended and walking technicolor.

Still, to obtain $S \simeq 0$, the relative values of the condensates
appearing in the OPE of the $V$\& $A$ correlators need to be altered with respect
to the QCD case. How this happens in a technicolor model is unclear to
us. On the
other hand, we have shown that the respective values of the condensates $o_V <
0$ and $o_V < o_A$ would follow from the simplest 5D modeling. This is quite
encouraging, and needs to be studied further, especially in connection with a
possible dual 4D description.

In this paper we described a scenario with fermions located on the UV brane,
although a natural setup would have bulk fermions, leading to a more
interesting phenomenology.

\acknowledgments

We thank Kaustubh Agashe, Tom Appelquist, Bogdan Dobrescu, Walter Goldberger,
Thomas Gr\'egoire, Ami Katz, Ken Lane, Ben Lillie, Adam Martin, Maurizio Piai,
Matt Schwartz, Witek Skiba, Raman Sundrum, Tim Tait, Jesse Thaler and the
Aspen Center for Physics. J.H. is supported by DOE grant
DE-FG02-92ER40704 and V.S. by grant DE-FG02-91ER40676.

\begin{appendix}
\section{From background fields to effective
metric}\label{appscalar}

Consider the action involving gravity, a set of $\tmop{SU} \left( N_f
\right)_L \times \tmop{SU} \left( N_f \right)_R$ Yang-Mills fields, a scalar
$X$ charged under the gauge symmetry as $(N_{f L}, N_{f R})$ and a neutral
scalar $\phi$,
\begin{eqnarray}
  S & = & \frac{1}{2 \kappa^2} \int d^5 x \sqrt{g}  \left( -\mathcal{R}-
  V_{\phi} + \frac{1}{2} g^{\tmop{MN}} \partial_M \phi \partial_N \phi \right)
  \nonumber\\
  &  & - \frac{1}{4 g_5^2}  \int \mathd^5 x \sqrt{g} g^{M N} g^{R S} 
  \left\langle L_{M R} L_{N S} + R_{M R} R_{N S} \right\rangle \nonumber\\
  &  & + \frac{1}{2 g_5^2}  \int \mathd^5 x \sqrt{g}  \left( g^{\tmop{MN}} 
  \left\langle D_M XD_N X \right\rangle - V_X \right)  \label{Sgrav2}
\end{eqnarray}
where $\left\langle \cdots \right\rangle$ means the trace in flavor space, and
$R_{M N} \equiv \partial_M R_N - \partial_N R_M - \mathi [R_M, R_N]$. The
square of the 5D YM coupling $g_5^2$ has dimensions of length. The action
(\ref{Sgrav2}) is invariant under ``parity'' $L \leftrightarrow R$ and the 5D
$\tmop{SU} \left( N_f \right) \times \tmop{SU} \left( N_f \right)$ gauge
transformations denoted by $R \left( x, z \right), L \left( x, z \right)$
acting as $R_M \equiv R_M^a T^a \longmapsto RR_M R^{\dag} + \mathi R
\partial_M R^{\dag}$. $\kappa$ is the 5D Newton constant related to the
curvature $l_0$ and the bulk cosmological constant by $1 / l_0^2 = - \kappa^2
\Lambda / 6$.

$X$ and $\phi$ gets vevs due to their potentials $V_{\phi, X}$, they fix the
vev profile in the fifth dimension and the BCs that these fields obey. The
condition for this potential to be {\tmem{natural}} is {\footnote{See
{\cite{DaRold:2005ju}} for an example of $V_X$.}}
\begin{eqnarray}
  V_X, V_{\phi} & \sim & \frac{1}{l_0^4},  \label{natV}
\end{eqnarray}
or, in other words, this potential generates vevs for $X$ and $\phi$ with BCs,
\begin{eqnarray}
  X (l_0), \phi (l_0) & \sim & 1 / l_0 . 
\end{eqnarray}
One cannot solve analytically the whole system. On the other hand, we are
interested in the net effect on the $\tmop{SU} (2)_L \times \tmop{SU} (2)_R$
fields. Therefore, one can solve for the system gravity+$\phi$ and add the
effect of the charged field $X$.

\subsection{Neutral scalar}\label{phi}

In this section we illustrate how the coupling of a scalar to gravity can
generate a metric that deviates from AdS as in Eq.(\ref{metric}). The scalar
does not break electroweak symmetry: its effect will be common to axial and
vector resonances.

We use the ansatz {\cite{DeWolfe:1999cp,Csaki:2000fc}}
\begin{eqnarray}
  \phi & = & \phi (z), 
\end{eqnarray}
to write down the equations of motion
\begin{eqnarray}
  \kappa^2 \phi'^2 & = & 6 \left( A - B \right) \\
  \kappa^2 V (\phi) & = & - \frac{3}{w^2} \left( A + B \right), 
\end{eqnarray}
where we have defined
\begin{eqnarray*}
  \begin{array}{l}
    A (z)
  \end{array} & = & 2 \frac{w'^2}{w^2}\\
  B (z) & = & \frac{w''}{w} .
\end{eqnarray*}
In AdS, $A = B = 2 / z^2$. In a metric of the form (\ref{metric2}),
\begin{eqnarray*}
  A & = & \frac{2}{z^2}  \left( 1 - \frac{4 o}{d - 1} \left( \frac{z}{l_1}
  \right)^{2 d} \right)\\
  B & = & \frac{2}{z^2}  \left( 1 + \frac{2 d - 3}{d - 1} o \left(
  \frac{z}{l_1} \right)^{2 d} \right) .
\end{eqnarray*}
The solution is
\begin{eqnarray}
  \phi (z) & = & \phi (l_0) + \frac{2}{d} \sqrt{\frac{- 3 (2 d + 1) o}{d - 1}}
  \left( \frac{z}{l_1} \right)^d \\
  V (\phi) & = & - \frac{12}{l_0^2} e^{\frac{d}{6 (2 d + 1)} (\phi -
  \phi_0)^2} \left( 1 - \frac{d^2}{24} \frac{2 d - 7}{2 d + 1} (\phi -
  \phi_0)^2 \right) 
\end{eqnarray}
The first thing we notice is that
\begin{eqnarray}
  \phi \tmop{non} - \tmop{tachyonic} & \Longrightarrow & o < 0 
  \label{nontach}
\end{eqnarray}
One can also check what will happen to gravity in this case. The graviton
equation of motion will receive a correction from the condensates that again
takes over the pure AdS near the IR:
\begin{eqnarray}
  - \psi^{''} + \frac{1}{z^2} \left( \frac{15}{4} + \delta V (z) \right) \psi
  (z) & = & m^2 \psi (z), 
\end{eqnarray}
where the extra piece in the potential is given by
\begin{eqnarray}
  \delta V (z) & = & 3 d \nu \left( \frac{z}{l_1} \right)^{2 d}  \left( 4 - 2
  d + 3 d \nu \left( \frac{z}{l_1} \right)^{2 d}  \right) . 
\end{eqnarray}
The zero mode is simply
\begin{eqnarray}
  \psi_0 (z) & = & \frac{w (z)^{3 / 2}}{N_0}, 
\end{eqnarray}
where $N_0$ is the norm of the graviton.

The only corrections to gravity will come from the tower of KK gravitons,
\begin{eqnarray}
  G_N & \sim & M_{5 D}^{- 3} \psi_0 (l_0)^2 . 
\end{eqnarray}
\subsection{Symmetry-breaking by a bulk scalar}\label{rewrite}

The $L R$ symmetry has to be broken near the IR brane. The standard way to model
this would be to introduce a bulk scalar that describes the lowest dimension
condensate associated with that breaking. Breaking by BCs on the other hand,
would only introduce non-local order parameters. Here we show that, for our
purposes, breaking by a bulk scalar is equivalent to introducing an effective
metric for the axial channel, and modifying the BCs.

If a bulk scalar $X$ transforming as a bifundamental under $\tmop{SU} \left(
N_f \right)_L \times \tmop{SU} \left( N_f \right)_R$ acquires a profile $v
(z)$, the wave equation for the axial KKs is modified as follows
\begin{eqnarray}
  - \frac{1}{w} \partial \left( w \partial \Phi \right) + 2 w^2 v^2 \Phi & = &
  M^2 \Phi,  \label{EOMw/scalar}
\end{eqnarray}
where the new term is the one proportional to $v^2$
{\cite{Erlich:2005qh,daRold:2005zs}}. The above equation can be recast in the
Schr\"odinger form
\begin{eqnarray}
  - \partial^2 \psi + V \psi & = & M^2 \psi,  \label{schro}
\end{eqnarray}
provided we define the new wave-functions $\psi$ as follows
\begin{eqnarray}
  \psi & = & \sqrt{w} \Phi .  \label{SLtoSch}
\end{eqnarray}
We can then read off the potential $V \left( z \right)$ in the Schr\"odinger
equation (\ref{schro}) as
\begin{eqnarray}
  V & = & \frac{\partial^2 \sqrt{w}}{\sqrt{w}} + w^2 v^2 . 
\end{eqnarray}
The point is now to invert the above trick of going from $\Phi$ to $\psi$, but
for a potential given by $V$ rather than simply by $\frac{\partial^2
\sqrt{w}}{\sqrt{w}}$. In other words, we want to solve for $w_X$ in the
second-order differential equation
\begin{eqnarray}
  \frac{\partial^2 \sqrt{w_X}}{\sqrt{w_X}} & = & V .  \label{diffeq}
\end{eqnarray}
The right solution can be picked by asking that the effective warp factor
$w_X$ be asymptotically AdS, i.e. the condition
\begin{eqnarray}
  \left. \frac{w_X}{w} \right|_{z = l_0} & \underset{l_0 \longrightarrow
  0}{\longrightarrow} & 1,  \label{wABC}
\end{eqnarray}
excludes the divergent linear combination, and also fixes the normalization.

The basic relation we need is then
\begin{eqnarray}
  \sqrt{w} \Phi & = & \sqrt{w_X} \varphi, 
\end{eqnarray}
so that the normalization condition is now
\begin{eqnarray}
  \int_{l_0}^{l_1} \mathd zw_X \varphi^2 & = & \mathcal{N}, 
\end{eqnarray}
and $(+)$ BCs for the original $\Phi$ wave-functions are modified into mixed
ones for the $\varphi$'s
\begin{eqnarray}
  - \partial \log \varphi & = & \frac{1}{2} \partial \log \frac{w_X}{w} . 
  \label{IRBCwv}
\end{eqnarray}
In the limit of a large condensate, $w_X$ deviates strongly from $w$ in the
IR, and this tends to a $(-)$ BC.

In an AdS background $w = l_0 / z$, the differential equation (\ref{diffeq})
reduces to
\begin{eqnarray}
  \frac{\partial^2 \sqrt{w_X}}{\sqrt{w_X}} & = & \frac{3}{4}  \frac{1}{z^2} +
  \left( \frac{l_0}{z} \right)^2 v \left( z \right)^2,  \label{diffeqAdS}
\end{eqnarray}
which can be solved analytically if $v \left( z \right)$ is a power-law.
\begin{eqnarray}
  v (z) & = & \sigma z^d 
\end{eqnarray}
\begin{eqnarray}
  \frac{w_X}{w} & = & _0 F_1 \left( ; \frac{d - 1}{d} ; \frac{\sigma^2 l_0^2
  z^{2 d}}{2 d^2} \right)^2  \underset{z \rightarrow 0}{\sim} 1 +
  \frac{\sigma^2 l_0^2}{d (d - 1)} z^{2 d} +\mathcal{O} \left( \frac{1}{d^4}
  \sigma^4 z^{4 d} \right) 
\end{eqnarray}
For the particular case of $d = 2$
\begin{eqnarray}
  d = 2 &  & w_X = \frac{l_0}{z} \tmop{Cosh}^2 \left( \frac{\sigma z^2}{2}
  \right) \underset{z \rightarrow 0}{\sim}  \frac{l_0}{z}  \left( 1 +
  \frac{\sigma^2}{4} z^4 + \frac{\sigma^4}{48} z^8 \ldots \right) 
\end{eqnarray}
Adding several fields of scaling dimensions $2, 3, 4 \ldots d$ would have an
effect on the metric suppressed by $(z / l_1)^{2 d}$. The lower the dimension,
the more the deviation from AdS extends inside the bulk. The effect is of
course maximum on the IR brane, but still there there is a suppression given
by the dimension that is suppressed by the dimension of the field,
\begin{eqnarray*}
  \left( \frac{1}{4}, \frac{1}{12}, \frac{1}{25}, \ldots \frac{1}{2 d (d - 1)}
  \right) &  & 
\end{eqnarray*}
In conclusion, for phenomenological purposes, one only needs to consider the
effect of the lower dimension condensates.

\section{Derivation of $S$ in any holographic model}\label{appA}

We derived a sum rule for $S$ in {\cite{Hirn:2005nr}}. It applies to 5D models
with BCs, where the symmetry breaking was limited to a crossed $L R$ term,
yielding different effective metrics for$V$and $A$. Here, we want to generalize
this result to the case where deviations from AdS (and in particular symmetry
breaking) are introduced by bulk scalars. For a quadratic quantity such as the
$S \tmop{parameter}, \tmop{this}$ can be done by using the results of Appendix
\ref{rewrite}. Indeed, there we showed how to rewrite the effect of bulk
scalars as effective metrics and effective BCs.

For any wave-equation with $(-)$ UV BC, we can write the decay constants of
the KK with wave-function $\varphi_n$ and mass $M_n$ as
\begin{eqnarray}
  \frac{g_5^2}{\sqrt{2}} f_n M_n^2 & = & \left. w \partial \varphi_n
  \right|_{l_0} .  \label{startf}
\end{eqnarray}
This is true whatever the basis, i.e. using $\Phi$, $\psi$ or $\varphi$
wave-functions. Indeed, whatever the representation used, the only
non-vanishing quadratic terms remaining after using the EOMs are surface
terms. Variation of these with respect to the source on the UV brane yields
their coupling to the KKs, i.e. the resonance decay constants.

To be general, we then consider mixed IR BCs in the form
\begin{eqnarray}
  \left. - \partial \log \varphi \right|_{l_1} & = & g_5^2 w \left( l_1
  \right)^2 M_{\tmop{IR}}^2 .  \label{mixedBC}
\end{eqnarray}
Eq.(\ref{startf}) can be recast as
\begin{eqnarray}
  \frac{g_5^2}{\sqrt{2}} f_n M_n^2 & = & \left. w \alpha^2 \partial \left(
  \frac{1}{\alpha} \varphi_n \right) \right|_{l_0},  \label{falpha}
\end{eqnarray}
provided we normalize the function $\alpha \left( z \right)$ such that
\begin{eqnarray}
  \alpha \left( l_0 \right) & = & 1 .  \label{UVBCalpha}
\end{eqnarray}
Other than that, the function $\alpha$ is undetermined at that stage. The
point is that, if $\alpha$ satisfies the same IR BC as $\varphi$
(\ref{mixedBC}), i.e.
\begin{eqnarray}
  \left. - \partial \log \alpha \right|_{l_1} & = & g_5^2 w \left( l_1
  \right)^2 M_{\tmop{IR}}^2,  \label{IRBCalpha}
\end{eqnarray}
and a massless spin-1 EOM
\begin{eqnarray}
  \partial \left( w \partial \alpha \right) & = & 0,  \label{EOMalpha}
\end{eqnarray}
then we can turn (\ref{falpha}) into a useful expression, namely
\begin{eqnarray}
  f_n & = & \frac{\sqrt{2}}{g_5^2}  \int_{l_0}^{l_1} \mathd z w \alpha
  \varphi_n . 
\end{eqnarray}

To make use of this, we still have to give the explicit expression for
$\alpha$
\begin{eqnarray}
  \alpha \left( z \right) & = & \frac{ \left( g_5^2 w \left( l_1 \right)^3
  M_{\tmop{IR}}^2 \right)^{- 1} + \int_z^{l_1} \frac{d z'}{w (z')}}{\left(
  g_5^2 w \left( l_1 \right)^3 M_{\tmop{IR}}^2 \right)^{- 1} +
  \int_{l_0}^{l_1} \frac{d z'}{w (z')}},  \label{gensolalpha}
\end{eqnarray}
which satisfies the EOM (\ref{EOMalpha}) and the two BCs (\ref{UVBCalpha}) and
(\ref{IRBCalpha}). It turns out that $\alpha$ can be interpreted in the
completely general case, as the wave-function of the GBs. This implies in
particular that the GB decay constant is given by
\begin{eqnarray}
  f^2 & = & \left. \frac{1}{g_{5^2}} w \partial \alpha \right|_{l_0} . 
  \label{f2}
\end{eqnarray}
For the model with SB implemented by BCs and two different metrics for $A$ and
$V$, we can check that the solutions for $\alpha$ were respectively
\begin{eqnarray}
  \alpha_V & \equiv & 1, \\
  \alpha_A & = & \frac{\int_z^{l_1} \frac{d z'}{w_A (z')}
  \bignone}{\int_{l_0}^{l_1} \frac{d z'}{w_A (z')} \bignone}, 
\end{eqnarray}
since the IR BCs for $V / A$ correspond to vanishing/infinite $M_{\tmop{IR}}$
respectively.

In the case where symmetry breaking is implemented by a bulk scalar, we can
rewrite the effect of the scalar vev $v$ on the axial wave-functions $\Phi_A$
as an effective metric $w_A$ felt by the wave-functions $\varphi_A$. We can
then simply apply the method of Section \ref{rewrite}, using $w_A$ as the
metric, and taking in to account the change of IR BC as follows.

For the transformed wave-functions $\varphi_A$, we obtain mixed BCs on the IR
brane, as indicated by (\ref{IRBCwv}). Indeed, (\ref{IRBCwv}) translates into
\begin{eqnarray}
  M_{\tmop{IR}}^2 & = & \left. \frac{1}{2 g_5^2 w_A^2} \partial \log
  \frac{w_A}{w_V} \right|_{l_1}, 
\end{eqnarray}
and we can then plug this into the solution for $\alpha$ (\ref{gensolalpha}).
Applying the completeness relation for the $\varphi_A$'s with the metric
$w_A$, we can derive
\begin{eqnarray}
  S & = & \frac{16 \mathpi}{g_5^2}  \int_{l_0}^{l_1} dz (w_V (z) \alpha_V
  (z)^2 - w_A (z) \alpha_A (z)^2) . 
\end{eqnarray}
This is the general result for $S$, valid using the expressions for $\alpha$
given in (\ref{gensolalpha}) for the case of mixed IR BCs. For the case with
bulk scalars, one needs to have determined beforehand the effective metric and
IR mass using the techniques of Section \ref{rewrite}.

\section{NDA for the condensates}\label{NDA}

We detail here the NDA estimates for the deviations from AdS. Whereas in
{\cite{Hirn:2005vk}} we estimated the natural size for condensates from the 4D
OPE, we propose here to start from the side of the 5D modeling.
Unsurprisingly, the results essentially agree, provided we account for missing
factors in {\cite{Hirn:2005vk}}, which do not matter for the low-dimension
condensates we are interested in.

Imagine the situation of Section \ref{rewrite}, i.e. a symmetry-breaking VEV
for the scalar field $X$. To perform dimensional analysis, we consider an AdS
background. Then, in order for this bulk profile to generate a dimension $2 d$
condensate in the axial two-point function, we have to assume that its 5D mass
is given by $m^2 l_0^2 = d \left( d - 4 \right)$. The scalar may then develop
a profile of the shape $v \left( z \right) = \frac{\sqrt{o}}{l_0}  \left(
\frac{z}{l_1} \right)^d$ where the IR value is set by a potential localized on
the IR brane. NDA on this potential (\ref{natV}) implies
\begin{eqnarray}
  o & = & \mathcal{O} \left( 1 \right) . 
\end{eqnarray}
This translates into a dimension $2 d$ condensate appearing in the two-point
function as
\begin{eqnarray}
  \Pi_A \left( Q^2 \right) & = & - \frac{N}{12 \mathpi^2} \log \left(
  \frac{Q^2}{\mu^2} \right) + \frac{\left\langle \mathcal{O}_{2 d}
  \right\rangle}{Q^{2 d}} + \ldots 
\end{eqnarray}
where {\cite{daRold:2005zs}}
\begin{eqnarray}
  \left\langle \mathcal{O}_{2 d} \right\rangle & = & \frac{1}{\sqrt{\pi}} 
  \frac{d}{d - 1}  \frac{\Gamma \left( d \right)^3}{\Gamma \left( d + 1 / 2
  \right)}  \frac{N}{12 \pi^2} ol_1^{- 2 d} .  \label{bigO}
\end{eqnarray}
Compared to the 4D estimate used in {\cite{Hirn:2005vk}}, this provides more
precise numerical factors. This includes a factorial growth with $d$ for $d
\gg 1$. It is interesting to note that the NDA analysis on the simple 5D model
produces the factorial growth expected in 4D {\cite{Shifman:1994yf}}. Note
that the factorial growth is expected for $d \gg 1$, but not necessarily for
$d \lesssim 3$: this is why the investigation in {\cite{Hirn:2005vk}} did not
include it, in order not to artificially enhance the effect on $S$. Even when
this growth is included, we see from Fig.\ref{fig-oVd} that a significant
effect on $S$ can be achieved only with a low-$d$ condensate.

What the factorial behavior is really telling us is that the OPE cannot be
resummed. Also, one may wonder whether adding higher and higher orders by
including additional scalars with a bulk profile is a convergent procedure. We
can answer this question by recasting the various profiles as a deformation of
the metric: this allows us to compare the respective deviations with the AdS
background. It turns out that the deviation from AdS is largest on the IR
brane, and goes down with $d$ for a profile $z^d / l_1^{d + 1}$ as
\begin{eqnarray}
  & \frac{1}{d \left( d - 1 \right)} . & 
\end{eqnarray}
Such a series can be summed, implying that the deformations from AdS, as
estimated from the model with scalars can be resummed, while still leading to
the (divergent) factorial growth in the OPE.

\section{Comparison 4D resonance saturation models}\label{res-sat}

In general, any Green's function becomes meromorphic in the large-$N$ limit
{\cite{Witten:1979kh}}. Thus, we can hope to model the two-point function by a
sum of poles, located at the masses of the resonances, and with residues
related to their decay constant. To get started, one may imagine a situation
where the vector and axial two-point spectral functions are equal above some
scale, so that $\tmop{Im} \Pi_{L R}$ vanishes above that scale. We then only
need to consider the {\tmem{finite number}} of resonances that are below that
scale. In that case, we have a finite number of parameters (the decay
constants and masses of the resonances). The result for the easiest cases are
as follows
\begin{itemize}
  \item Only one resonance: the first WSR fixes it to be a vector, which
  implies $S > 0$.
  
  \item Two resonances: the first WSR fixes one of them to be a vector.
  Getting $S \leqslant 0$ requires the lightest resonance to be axial. Using
  the second WSR, one would then find $\left\langle \mathcal{O}_4
  \right\rangle_{V - A} > 0$, in conflict with Witten's positivity constraint.
  
  \item Three resonances: assuming $\left\langle \mathcal{O}_4
  \right\rangle_{V - A} = 0$ as in QCD, the authors of {\cite{Knecht:1997ts}}
  have shown that it its possible to get $S \leqslant 0$ without encountering
  any of the above-mentioned problems (i.e. they have $\left\langle
  \mathcal{O}_6 \right\rangle_{V - A} < 0$). The spectrum they find is then:  $A V A$.
\end{itemize}
In such modeling, there is in fact a conflict between obtaining $S \leqslant
0$ and satisfying Witten's positivity constraint for any even number of
resonances. This is really a problem of the model itself, since in reality
there should be an infinite number of resonances for large-$N$
{\footnote{Alternatively, for finite $N$, the resonances should get a finite
width and the spectral function modified accordingly.}}.

However, the answer in such 4D models only depends on the $V - A$ condensate,
not on the two condensates separately, whereas we've seen in the 5D model that
the answer depended on both (see Fig.\ref{fig-S/Ncontour}). Whereas a 5D model
relates $f$ with the axial condensates, as should be, in 4D models, it is an
input parameter. In the 5D case, the cancellation of the pion pole against
resonance contributions to yield a vanishing dimension-2 axial condensate is
built in. The dimension-2 condensate thus automatically vanishes unless it is
explicitly included in the model. This is not automatic in the 4D resonance
saturation approach.

In addition to these concerns, we point out that the 5D model predicts the
resonance couplings to the fermion currents, depending on the localization of
the latter. In a generic 4D model, these couplings would be arbitrary.

To summarize, 4D models of resonances work in the following way: starting with
the input of $o_{V - A}, \frac{M_{\rho}}{f}$ and $\frac{M_{a_1}}{f}$ one uses
WSRs to compute $S, f_{a_1}$ and $f_{\rho}$. Schematically,
\begin{eqnarray*}
  \text{4D model} : \mathcal{O}_{V - A}, \frac{M_{\rho}}{f}, \frac{M_{a_1}}{f}
  & \underset{\tmop{WSRs}}{\Longrightarrow} & S, f_{\rho}, f_{a_1} 
\end{eqnarray*}
On the other hand, the 5D model works differently,
\begin{eqnarray*}
  \text{5D model} : N, o_V, o_A & \Longrightarrow &  \frac{M_{\rho}}{f},
  \frac{M_{a_1}}{f}, S, f_{\rho}, f_{a_1} 
\end{eqnarray*}
There are, of course, other advantages in using a 5D model besides parameter
counting: the correspondence between 4D quantities and 5D objects is very
intuitive (see Sec.\ref{analogueintro}).

\section{Link with TC scales}\label{TC}

We try to relate the present results with previous ideas about the behavior of
strong 4D theories. We explain this on the example of walking technicolor, in
which case the high-energy and intermediate-energy scalings of the
techni-quark condensate are different. To be specific, the OPE of $\Pi_{V, A}$
should include a dimension-6 condensate for $Q^2 \longrightarrow + \infty$
\begin{eqnarray}
  \Pi_X \left( Q^2 \right) & \underset{Q^2 > \Lambda_{\ast}}{=} & -
  \frac{N}{12 \mathpi^2} \log \left( \frac{Q^2}{\mu^2} \right) +
  \frac{\left\langle \mathcal{O}_6 \right\rangle}{Q^6} + \ldots 
\end{eqnarray}
At energies below the critical scale $\Lambda_{\ast}$ where the
coupling constant walks, a large anomalous dimension may be generated for the
techni-quark condensate
{\cite{Appelquist:1986an,Yamawaki:1985zg,Holdom:1981rm}} for the extreme
walking case, yielding
\begin{eqnarray}
  \Pi_X \left( Q^2 \right) & \underset{\Lambda_{\ast} > Q^2 >
  \Lambda_{\tmop{TC}}}{\simeq} & - \frac{N}{12 \mathpi^2} \log \left(
  \frac{Q^2}{\mu^2} \right) + \frac{\left\langle \mathcal{O}_4
  \right\rangle}{Q^4} + \ldots  \label{appOPE}
\end{eqnarray}
for scales much larger than the confinement scale $\Lambda_{\tmop{TC}}$, but
smaller than $\Lambda_{\ast}$. 
The behavior (\ref{appOPE}) is the one that has to be reproduced by the model,
since it is the one which influences the value of $S$. Translating into 5D
requires the identification of the confinement scale
\begin{eqnarray}
  \Lambda_{\tmop{TC}} & \sim & 1 / l_1 . 
\end{eqnarray}
For the high scales, the exact translation will be model-dependent. One
expects the localization of the fermion to correspond to the inverse of the
scale at which they get their masses (assuming that it comes from an order one
5D coupling). Sticking to the simplest model with fermions on the UV brane,
that scale is of order $1 / l_0$. This would correspond to the extended
technicolor scale $\Lambda_{\tmop{ETC}}$ if one was thinking of modeling such
a 4D set-up, in which case there should be more than just two flavors of
techni-quarks, and one must discuss the issue of explicit breaking to lift the
physical pseudo-GBs above the experimental limits {\cite{Eichten:1979ah}}. We
do not consider such a scenario here, but it is still useful to keep in mind
the correspondence
\begin{eqnarray}
  \Lambda_{\tmop{ETC}} & \sim & 1 / l_0 . 
\end{eqnarray}
One comment is in order about the respective sizes of $\Lambda_{\tmop{ETC}}$
and $\Lambda_{\ast}$. It looks as if we made the hidden assumption
$\Lambda_{\ast} > \Lambda_{\tmop{ETC}}$, since we have used the extreme
walking approximation (\ref{appOPE}) up to the scale $\Lambda_{\tmop{ETC}}$.
However, since the $S$ parameter is a UV-independent quantity, the ordering of
the two scales $\Lambda_{\tmop{TC}}$ and $\Lambda_{\ast}$ is irrelevant to the
present discussion: having the switch-over from the $1 / Q^6$ ($z^6$) behavior
to the $1 / Q^4$ ($z^4$) behavior above or below $\Lambda_{\tmop{ETC}}$
(before or after $1 / l_0$) is numerically unimportant. In Fig.\ref{fig-Sz},
only the contributions from $S \left( z \right)$ near the UV brane would be
affected, and they would remain small.
\end{appendix}

\bibliography{analogue}

\providecommand{\href}[2]{#2}\begingroup\raggedright\begin{thebibliography}{10}

\bibitem{Maldacena:1997re}
J.~M. Maldacena, {\it The large n limit of superconformal field theories and
  supergravity},  {\em Adv. Theor. Math. Phys.} {\bf 2} (1998) 231--252,
  [\href{http://xxx.lanl.gov/abs/hep-th/9711200}{{\tt hep-th/9711200}}].

\bibitem{Gubser:1998bc}
S.~S. Gubser, I.~R. Klebanov, and A.~M. Polyakov, {\it Gauge theory correlators
  from non-critical string theory},  {\em Phys. Lett.} {\bf B428} (1998)
  105--114, [\href{http://xxx.lanl.gov/abs/hep-th/9802109}{{\tt
  hep-th/9802109}}].

\bibitem{Witten:1998qj}
E.~Witten, {\it Anti-de sitter space and holography},  {\em Adv. Theor. Math.
  Phys.} {\bf 2} (1998) 253--291,
  [\href{http://xxx.lanl.gov/abs/hep-th/9802150}{{\tt hep-th/9802150}}].

\bibitem{Arkani-Hamed:2000ds}
N.~Arkani-Hamed, M.~Porrati, and L.~Randall, {\it Holography and
  phenomenology},  {\em JHEP} {\bf 08} (2001) 017,
  [\href{http://xxx.lanl.gov/abs/hep-th/0012148}{{\tt hep-th/0012148}}].

\bibitem{Pomarol:2000hp}
A.~Pomarol, {\it Grand unified theories without the desert},  {\em Phys. Rev.
  Lett.} {\bf 85} (2000) 4004--4007,
  [\href{http://xxx.lanl.gov/abs/hep-ph/0005293}{{\tt hep-ph/0005293}}].

\bibitem{Erlich:2005qh}
J.~Erlich, E.~Katz, D.~T. Son, and M.~A. Stephanov, {\it Qcd and a holographic
  model of hadrons},  {\em Phys. Rev. Lett.} {\bf 95} (2005) 261602,
  [\href{http://xxx.lanl.gov/abs/hep-ph/0501128}{{\tt hep-ph/0501128}}].

\bibitem{daRold:2005zs}
L.~Da~Rold and A.~Pomarol, {\it Chiral symmetry breaking from five dimensional
  spaces},  {\em Nucl. Phys.} {\bf B721} (2005) 79--97,
  [\href{http://xxx.lanl.gov/abs/hep-ph/0501218}{{\tt hep-ph/0501218}}].

\bibitem{Katz:2005ir}
E.~Katz, A.~Lewandowski, and M.~D. Schwartz, {\it Tensor mesons in ads/qcd},
  {\em Phys. Rev.} {\bf D74} (2006) 086004,
  [\href{http://xxx.lanl.gov/abs/hep-ph/0510388}{{\tt hep-ph/0510388}}].

\bibitem{Sikivie:1980hm}
P.~Sikivie, L.~Susskind, M.~B. Voloshin, and V.~I. Zakharov, {\it Isospin
  breaking in technicolor models},  {\em Nucl. Phys.} {\bf B173} (1980) 189.

\bibitem{Agashe:2006at}
K.~Agashe, R.~Contino, L.~Da~Rold, and A.~Pomarol, {\it A custodial symmetry
  for z b anti-b},  {\em Phys. Lett.} {\bf B641} (2006) 62--66,
  [\href{http://xxx.lanl.gov/abs/hep-ph/0605341}{{\tt hep-ph/0605341}}].

\bibitem{Weinberg:1979bn}
S.~Weinberg, {\it Implications of dynamical symmetry breaking: An addendum},
  {\em Phys. Rev.} {\bf D19} (1979) 1277--1280.

\bibitem{Susskind:1978ms}
L.~Susskind, {\it Dynamics of spontaneous symmetry breaking in the weinberg-
  salam theory},  {\em Phys. Rev.} {\bf D20} (1979) 2619--2625.

\bibitem{Eichten:1979ah}
E.~Eichten and K.~D. Lane, {\it Dynamical breaking of weak interaction
  symmetries},  {\em Phys. Lett.} {\bf B90} (1980) 125--130.

\bibitem{Dimopoulos:1979es}
S.~Dimopoulos and L.~Susskind, {\it Mass without scalars},  {\em Nucl. Phys.}
  {\bf B155} (1979) 237--252.

\bibitem{Peskin:1991sw}
M.~E. Peskin and T.~Takeuchi, {\it Estimation of oblique electroweak
  corrections},  {\em Phys. Rev.} {\bf D46} (1992) 381--409.

\bibitem{hep-ph/0306259}
R.~Contino, Y.~Nomura, and A.~Pomarol, {\it Higgs as a holographic
  pseudo-goldstone boson},  {\em Nucl. Phys.} {\bf B671} (2003) 148--174,
  [\href{http://xxx.lanl.gov/abs/hep-ph/0306259}{{\tt hep-ph/0306259}}].

\bibitem{Agashe:2004rs}
K.~Agashe, R.~Contino, and A.~Pomarol, {\it The minimal composite higgs model},
   {\em Nucl. Phys.} {\bf B719} (2005) 165--187,
  [\href{http://xxx.lanl.gov/abs/hep-ph/0412089}{{\tt hep-ph/0412089}}].

\bibitem{Contino:2006qr}
R.~Contino, L.~Da~Rold, and A.~Pomarol, {\it Light custodians in natural
  composite higgs models},  \href{http://xxx.lanl.gov/abs/hep-ph/0612048}{{\tt
  hep-ph/0612048}}.

\bibitem{Barbieri:2004qk}
R.~Barbieri, A.~Pomarol, R.~Rattazzi, and A.~Strumia, {\it Electroweak symmetry
  breaking after lep1 and lep2},  {\em Nucl. Phys.} {\bf B703} (2004) 127--146,
  [\href{http://xxx.lanl.gov/abs/hep-ph/0405040}{{\tt hep-ph/0405040}}].

\bibitem{hep-ph/0604111}
G.~Cacciapaglia, C.~Csaki, G.~Marandella, and A.~Strumia, {\it The minimal set
  of electroweak precision parameters},  {\em Phys. Rev.} {\bf D74} (2006)
  033011, [\href{http://xxx.lanl.gov/abs/hep-ph/0604111}{{\tt
  hep-ph/0604111}}].

\bibitem{Yao:2006px}
{\bf Particle Data Group} Collaboration, W.~M. Yao {\em et~al.}, {\it Review of
  particle physics},  {\em J. Phys.} {\bf G33} (2006) 1--1232.

\bibitem{Holdom:1990tc}
B.~Holdom and J.~Terning, {\it Large corrections to electroweak parameters in
  technicolor theories},  {\em Phys. Lett.} {\bf B247} (1990) 88--92.

\bibitem{Luty:2004ye}
M.~A. Luty and T.~Okui, {\it Conformal technicolor},  {\em JHEP} {\bf 09}
  (2006) 070, [\href{http://xxx.lanl.gov/abs/hep-ph/0409274}{{\tt
  hep-ph/0409274}}].

\bibitem{Barbieri:2003pr}
R.~Barbieri, A.~Pomarol, and R.~Rattazzi, {\it Weakly coupled higgsless
  theories and precision electroweak tests},  {\em Phys. Lett.} {\bf B591}
  (2004) 141--149, [\href{http://xxx.lanl.gov/abs/hep-ph/0310285}{{\tt
  hep-ph/0310285}}].

\bibitem{Csaki:2003zu}
C.~Csaki, C.~Grojean, L.~Pilo, and J.~Terning, {\it Towards a realistic model
  of higgsless electroweak symmetry breaking},  {\em Phys. Rev. Lett.} {\bf 92}
  (2004) 101802, [\href{http://xxx.lanl.gov/abs/hep-ph/0308038}{{\tt
  hep-ph/0308038}}].

\bibitem{hep-ph/0309189}
Y.~Nomura, {\it Higgsless theory of electroweak symmetry breaking from warped
  space},  {\em JHEP} {\bf 11} (2003) 050,
  [\href{http://xxx.lanl.gov/abs/hep-ph/0309189}{{\tt hep-ph/0309189}}].

\bibitem{Agashe:2003zs}
K.~Agashe, A.~Delgado, M.~J. May, and R.~Sundrum, {\it Rs1, custodial isospin
  and precision tests},  {\em JHEP} {\bf 08} (2003) 050,
  [\href{http://xxx.lanl.gov/abs/hep-ph/0308036}{{\tt hep-ph/0308036}}].

\bibitem{Son:2003et}
D.~T. Son and M.~A. Stephanov, {\it Qcd and dimensional deconstruction},  {\em
  Phys. Rev.} {\bf D69} (2004) 065020,
  [\href{http://xxx.lanl.gov/abs/hep-ph/0304182}{{\tt hep-ph/0304182}}].

\bibitem{Hirn:2005vk}
J.~Hirn, N.~Rius, and V.~Sanz, {\it Geometric approach to condensates in
  holographic qcd},  {\em Phys. Rev.} {\bf D73} (2006) 085005,
  [\href{http://xxx.lanl.gov/abs/hep-ph/0512240}{{\tt hep-ph/0512240}}].

\bibitem{Hirn:2006nt}
J.~Hirn and V.~Sanz, {\it A negative s parameter from holographic technicolor},
   {\em Phys. Rev. Lett.} {\bf 97} (2006) 121803,
  [\href{http://xxx.lanl.gov/abs/hep-ph/0606086}{{\tt hep-ph/0606086}}].

\bibitem{Cacciapaglia:2006mz}
G.~Cacciapaglia, C.~Csaki, G.~Marandella, and J.~Terning, {\it The gaugephobic
  higgs},  \href{http://xxx.lanl.gov/abs/hep-ph/0611358}{{\tt hep-ph/0611358}}.

\bibitem{Cacciapaglia:2004jz}
G.~Cacciapaglia, C.~Csaki, C.~Grojean, and J.~Terning, {\it Oblique corrections
  from higgsless models in warped space},  {\em Phys. Rev.} {\bf D70} (2004)
  075014, [\href{http://xxx.lanl.gov/abs/hep-ph/0401160}{{\tt
  hep-ph/0401160}}].

\bibitem{Manohar:1983md}
A.~Manohar and H.~Georgi, {\it Chiral quarks and the nonrelativistic quark
  model},  {\em Nucl. Phys.} {\bf B234} (1984) 189.

\bibitem{Georgi:1986kr}
H.~Georgi and L.~Randall, {\it Flavor conserving cp violation in invisible
  axion models},  {\em Nucl. Phys.} {\bf B276} (1986) 241.

\bibitem{Luty:1997fk}
M.~A. Luty, {\it Naive dimensional analysis and supersymmetry},  {\em Phys.
  Rev.} {\bf D57} (1998) 1531--1538,
  [\href{http://xxx.lanl.gov/abs/hep-ph/9706235}{{\tt hep-ph/9706235}}].

\bibitem{Cohen:1997rt}
A.~G. Cohen, D.~B. Kaplan, and A.~E. Nelson, {\it Counting 4pi's in strongly
  coupled supersymmetry},  {\em Phys. Lett.} {\bf B412} (1997) 301--308,
  [\href{http://xxx.lanl.gov/abs/hep-ph/9706275}{{\tt hep-ph/9706275}}].

\bibitem{Chacko:1999hg}
Z.~Chacko, M.~A. Luty, and E.~Ponton, {\it Massive higher-dimensional gauge
  fields as messengers of supersymmetry breaking},  {\em JHEP} {\bf 07} (2000)
  036, [\href{http://xxx.lanl.gov/abs/hep-ph/9909248}{{\tt hep-ph/9909248}}].

\bibitem{Randall:2001gc}
L.~Randall and M.~D. Schwartz, {\it Unification and the hierarchy from ads5},
  {\em Phys. Rev. Lett.} {\bf 88} (2002) 081801,
  [\href{http://xxx.lanl.gov/abs/hep-th/0108115}{{\tt hep-th/0108115}}].

\bibitem{Randall:2002tg}
L.~Randall, V.~Sanz, and M.~D. Schwartz, {\it Entropy-area relations in field
  theory},  {\em JHEP} {\bf 06} (2002) 008,
  [\href{http://xxx.lanl.gov/abs/hep-th/0204038}{{\tt hep-th/0204038}}].

\bibitem{hep-ph/0111016}
R.~Sekhar~Chivukula, D.~A. Dicus, and H.-J. He, {\it Unitarity of compactified
  five dimensional yang-mills theory},  {\em Phys. Lett.} {\bf B525} (2002)
  175--182, [\href{http://xxx.lanl.gov/abs/hep-ph/0111016}{{\tt
  hep-ph/0111016}}].

\bibitem{Chivukula:2002ej}
R.~S. Chivukula and H.-J. He, {\it Unitarity of deconstructed five-dimensional
  yang-mills theory},  {\em Phys. Lett.} {\bf B532} (2002) 121--128,
  [\href{http://xxx.lanl.gov/abs/hep-ph/0201164}{{\tt hep-ph/0201164}}].

\bibitem{DaRold:2005ju}
L.~Da~Rold and A.~Pomarol, {\it Chiral symmetry breaking from five dimensional
  spaces},  {\em PoS} {\bf HEP2005} (2006) 355.

\bibitem{DaRold:2005vr}
L.~Da~Rold and A.~Pomarol, {\it The scalar and pseudoscalar sector in a
  five-dimensional approach to chiral symmetry breaking},  {\em JHEP} {\bf 01}
  (2006) 157, [\href{http://xxx.lanl.gov/abs/hep-ph/0510268}{{\tt
  hep-ph/0510268}}].

\bibitem{Csaki:2006ji}
C.~Csaki and M.~Reece, {\it Toward a systematic holographic qcd: A braneless
  approach},  \href{http://xxx.lanl.gov/abs/hep-ph/0608266}{{\tt
  hep-ph/0608266}}.

\bibitem{Witten:1983ut}
E.~Witten, {\it Some inequalities among hadron masses},  {\em Phys. Rev. Lett.}
  {\bf 51} (1983) 2351.

\bibitem{Piai:2006hy}
M.~Piai, {\it Precision electro-weak parameters from ads(5), localized kinetic
  terms and anomalous dimensions},
  \href{http://xxx.lanl.gov/abs/hep-ph/0608241}{{\tt hep-ph/0608241}}.

\bibitem{Csaki:2002gy}
C.~Csaki, J.~Erlich, and J.~Terning, {\it The effective lagrangian in the
  randall-sundrum model and electroweak physics},  {\em Phys. Rev.} {\bf D66}
  (2002) 064021, [\href{http://xxx.lanl.gov/abs/hep-ph/0203034}{{\tt
  hep-ph/0203034}}].

\bibitem{Goldberger:1999uk}
W.~D. Goldberger and M.~B. Wise, {\it Modulus stabilization with bulk fields},
  {\em Phys. Rev. Lett.} {\bf 83} (1999) 4922--4925,
  [\href{http://xxx.lanl.gov/abs/hep-ph/9907447}{{\tt hep-ph/9907447}}].

\bibitem{Goldberger:1999un}
W.~D. Goldberger and M.~B. Wise, {\it Phenomenology of a stabilized modulus},
  {\em Phys. Lett.} {\bf B475} (2000) 275--279,
  [\href{http://xxx.lanl.gov/abs/hep-ph/9911457}{{\tt hep-ph/9911457}}].

\bibitem{hep-ph/9502247}
R.~Casalbuoni {\em et~al.}, {\it Symmetries for vector and axial vector
  mesons},  {\em Phys. Lett.} {\bf B349} (1995) 533--540,
  [\href{http://xxx.lanl.gov/abs/hep-ph/9502247}{{\tt hep-ph/9502247}}].

\bibitem{Appelquist:1999dq}
T.~Appelquist, P.~S. Rodrigues~da Silva, and F.~Sannino, {\it Enhanced global
  symmetries and the chiral phase transition},  {\em Phys. Rev.} {\bf D60}
  (1999) 116007, [\href{http://xxx.lanl.gov/abs/hep-ph/9906555}{{\tt
  hep-ph/9906555}}].

\bibitem{hep-ph/0405188}
R.~Casalbuoni, S.~De~Curtis, and D.~Dominici, {\it Moose models with vanishing
  s parameter},  {\em Phys. Rev.} {\bf D70} (2004) 055010,
  [\href{http://xxx.lanl.gov/abs/hep-ph/0405188}{{\tt hep-ph/0405188}}].

\bibitem{Abe:1997fd}
{\bf CDF} Collaboration, F.~Abe {\em et~al.}, {\it Search for new gauge bosons
  decaying into dileptons in $\bar{p}p$ collisions at $\sqrt{s} = 1.8$ tev},
  {\em Phys. Rev. Lett.} {\bf 79} (1997) 2192--2197.

\bibitem{Karch:2006pv}
A.~Karch, E.~Katz, D.~T. Son, and M.~A. Stephanov, {\it Linear confinement and
  ads/qcd},  {\em Phys. Rev.} {\bf D74} (2006) 015005,
  [\href{http://xxx.lanl.gov/abs/hep-ph/0602229}{{\tt hep-ph/0602229}}].

\bibitem{Lane:1993wz}
K.~D. Lane, {\it An introduction to technicolor},
  \href{http://xxx.lanl.gov/abs/hep-ph/9401324}{{\tt hep-ph/9401324}}.

\bibitem{Appelquist:1998xf}
T.~Appelquist and F.~Sannino, {\it The physical spectrum of conformal su(n)
  gauge theories},  {\em Phys. Rev.} {\bf D59} (1999) 067702,
  [\href{http://xxx.lanl.gov/abs/hep-ph/9806409}{{\tt hep-ph/9806409}}].

\bibitem{Sundrum:1991rf}
R.~Sundrum and S.~D.~H. Hsu, {\it Walking technicolor and electroweak radiative
  corrections},  {\em Nucl. Phys.} {\bf B391} (1993) 127--146,
  [\href{http://xxx.lanl.gov/abs/hep-ph/9206225}{{\tt hep-ph/9206225}}].

\bibitem{DeWolfe:1999cp}
O.~DeWolfe, D.~Z. Freedman, S.~S. Gubser, and A.~Karch, {\it Modeling the fifth
  dimension with scalars and gravity},  {\em Phys. Rev.} {\bf D62} (2000)
  046008, [\href{http://xxx.lanl.gov/abs/hep-th/9909134}{{\tt
  hep-th/9909134}}].

\bibitem{Csaki:2000fc}
C.~Csaki, J.~Erlich, T.~J. Hollowood, and Y.~Shirman, {\it Universal aspects of
  gravity localized on thick branes},  {\em Nucl. Phys.} {\bf B581} (2000)
  309--338, [\href{http://xxx.lanl.gov/abs/hep-th/0001033}{{\tt
  hep-th/0001033}}].

\bibitem{Hirn:2005nr}
J.~Hirn and V.~Sanz, {\it Interpolating between low and high energy qcd via a
  5d yang-mills model},  {\em JHEP} {\bf 12} (2005) 030,
  [\href{http://xxx.lanl.gov/abs/hep-ph/0507049}{{\tt hep-ph/0507049}}].

\bibitem{Shifman:1994yf}
M.~A. Shifman, {\it Theory of pre-asymptotic effects in weak inclusive decays},
   \href{http://xxx.lanl.gov/abs/hep-ph/9405246}{{\tt hep-ph/9405246}}.

\bibitem{Witten:1979kh}
E.~Witten, {\it Baryons in the 1/n expansion},  {\em Nucl. Phys.} {\bf B160}
  (1979) 57.

\bibitem{Knecht:1997ts}
M.~Knecht and E.~de~Rafael, {\it Patterns of spontaneous chiral symmetry
  breaking in the large n(c) limit of qcd-like theories},  {\em Phys. Lett.}
  {\bf B424} (1998) 335--342,
  [\href{http://xxx.lanl.gov/abs/hep-ph/9712457}{{\tt hep-ph/9712457}}].

\bibitem{Appelquist:1986an}
T.~W. Appelquist, D.~Karabali, and L.~C.~R. Wijewardhana, {\it Chiral
  hierarchies and the flavor changing neutral current problem in technicolor},
  {\em Phys. Rev. Lett.} {\bf 57} (1986) 957.

\bibitem{Yamawaki:1985zg}
K.~Yamawaki, M.~Bando, and K.-i. Matumoto, {\it Scale invariant technicolor
  model and a technidilaton},  {\em Phys. Rev. Lett.} {\bf 56} (1986) 1335.

\bibitem{Holdom:1981rm}
B.~Holdom, {\it Raising the sideways scale},  {\em Phys. Rev.} {\bf D24} (1981)
  1441.

\end{thebibliography}\endgroup

\bibliographystyle{jhep}

\end{document}